\documentclass[aps,twocolumn,prd,superscriptaddress,preprintnumbers,showpacs]{revtex4-1}
\usepackage{graphicx}
\usepackage{bm}
\usepackage{amsmath}
\usepackage{amssymb}
\usepackage{amsthm}

\newcommand{\be}{\begin{equation}}
\newcommand{\ee}{\end{equation}}
\newcommand{\bea}{\begin{eqnarray}}
\newcommand{\eea}{\end{eqnarray}}
\newcommand{\bdm}{\begin{displaymath}}
\newcommand{\edm}{\end{displaymath}}
\newcommand{\beas}{\begin{eqnarray*}}
\newcommand{\eeas}{\end{eqnarray*}}

\begin{document}
\title{Current and future constraints on extended Bekenstein-type models for a varying fine-structure constant}
\author{C. S. Alves}
\email[]{catarina.alves17@imperial.ac.uk}
\affiliation{Centro de Astrof\'{\i}sica, Universidade do Porto, Rua das Estrelas, 4150-762 Porto, Portugal}
\affiliation{Instituto de Astrof\'{\i}sica e Ci\^encias do Espa\c co, CAUP, Rua das Estrelas, 4150-762 Porto, Portugal}
\affiliation{Department of Mathematics, Imperial College London, London SW7 2AZ, United Kingdom}
\author{A. C. O. Leite}
\email[]{Ana.Leite@astro.up.pt}
\affiliation{Centro de Astrof\'{\i}sica, Universidade do Porto, Rua das Estrelas, 4150-762 Porto, Portugal}
\affiliation{Instituto de Astrof\'{\i}sica e Ci\^encias do Espa\c co, CAUP, Rua das Estrelas, 4150-762 Porto, Portugal}
\affiliation{Faculdade de Ci\^encias, Universidade do Porto, Rua do Campo Alegre, 4150-007 Porto, Portugal}
\author{C. J. A. P. Martins}
\email[]{Carlos.Martins@astro.up.pt}
\affiliation{Centro de Astrof\'{\i}sica, Universidade do Porto, Rua das Estrelas, 4150-762 Porto, Portugal}
\affiliation{Instituto de Astrof\'{\i}sica e Ci\^encias do Espa\c co, CAUP, Rua das Estrelas, 4150-762 Porto, Portugal}
\author{T. A. Silva}
\email[]{up201405824@fc.up.pt}
\affiliation{Centro de Astrof\'{\i}sica, Universidade do Porto, Rua das Estrelas, 4150-762 Porto, Portugal}
\affiliation{Faculdade de Ci\^encias, Universidade do Porto, Rua do Campo Alegre, 4150-007 Porto, Portugal}
\author{S. A. Berge}
\email[]{uay1592@edu.kunskapsgymnasiet.se}
\affiliation{Internationella kunskapsgymnasiet, Buteljgatan 4, 117 43 Stockholm, Sweden}
\author{B. S. A. Silva}
\email[]{beatriz.ataide.silva@tecnico.ulisboa.pt}
\affiliation{Instituto Superior T\'ecnico, Universidade de Lisboa, Avenida Rovisco Pais, 1049-001 Lisboa, Portugal}

\date{4 December 2017}

\begin{abstract}
There is a growing interest in astrophysical tests of the stability of dimensionless fundamental couplings, such as the fine-structure constant $\alpha$, as an optimal probe of new physics. The imminent arrival of the ESPRESSO spectrograph will soon enable significant gains in the precision and accuracy of these tests and widen the range of theoretical models that can be tightly constrained. Here we illustrate this by studying proposed extensions of the Bekenstein-type models for the evolution of $\alpha$ that allow different couplings of the scalar field to both dark matter and dark energy. We use a combination of current astrophysical and local laboratory data (from tests with atomic clocks) to show that these couplings are constrained to parts per million level, with the constraints being dominated by the atomic clocks. We also quantify the expected improvements from ESPRESSO and other future spectrographs, and briefly discuss possible observational strategies, showing that these facilities can improve current constraints by more than an order of magnitude.
\end{abstract}
\maketitle

\section{\label{intro}Introduction} 

Astrophysical tests of the stability of fundamental couplings such as the fine-structure constant $\alpha$ are an extremely active area of observational research. The deep and compelling conceptual importance of carrying out these tests has been complemented by recent (although controversial) evidence for such a variation \cite{Webb}, coming from high-resolution optical/UV spectroscopic measurements of absorption systems along the line of sight of bright quasars, and by the growing realization that even null results at the best currently available levels of sensitivity already tightly constrain many cosmology and particle physics paradigms. A recent review of the field can be found in \cite{ROPP}.

Arguably the simplest class of phenomenological models for varying $\alpha$ is the one first suggested by Bekenstein \cite{Bekenstein} where, by construction, the dynamical degree of freedom responsible for this variation has a negligible effect on the cosmological dynamics. These models are characterized by a single dimensionless phenomenological parameter, which describes the strength of the coupling of the dynamical scalar degree of freedom to the electromagnetic sector, and therefore also determines the amount of Weak Equivalence Principle violation in these models---see \cite{SBM,Carroll,Dvali,Chiba} for a discussion of the relation between the two. The cosmological implications of these models were first explored by Sandvik, Barrow and Magueijo \cite{SBM}, who also obtained some qualitative constraints on the model. Stronger constraints, benefiting both from additional data and from a more detailed statistical analysis, were later obtained in \cite{Leal,LeiteBek}.

In what follows we extend these earlier works by studying a broader class of Bekenstein-type models discussed by Olive and Pospelov \cite{OlivePospelov}. In this case different couplings to the dark matter and dark energy sectors are allowed, and the behavior of $\alpha$ depends on both of them. This would immediately suggest that the two parameters will be degenerate, but as we show a combination of the available astrophysical measurements of $\alpha$, which thus far have been carried out in the approximate redshift range $0<z<4$, and local laboratory tests (plus, optionally, cosmological background data) partially breaks this degeneracy and leads to strong constraints on both parameters.

Part of our motivation for this work stems from the fact that more precise and accurate tests of the stability of $\alpha$ will be available in the near future. Indeed, improving these tests is a flagship science case for the ESPRESSO spectrograph \cite{ESPRESSO}---whose commissioning is ongoing at the time of writing---as well as for next-generation instruments such as ELT-HIRES \cite{HIRES} (and analogous instruments at other extremely large telescopes). Meanwhile, the sensitivity of local tests with atomic clocks is also expected to improve. Bearing this in mind---and noting that forecasts for this class of models can be done in all generality using Fisher Matrix analysis techniques---we also study how current constraints can be improved in the coming years, focusing on the case of ESPRESSO (whose technical specifications are all known) but also providing a more general discussion that may apply to future spectrographs.

The plan for the rest of the paper is as follows. In Sect. \ref{extbek} we very briefly introduce Bekenstein's varying $\alpha$ model, and then discuss its extension by Olive and Pospelov. Current constraints on this model are then obtained in Sect. \ref{currdata}, while in Sect. \ref{next} we discuss forecasts for future facilities, with some emphasis being given to the ESPRESSO spectrograph but also considering more general choices of numbers, sensitivities and redshift distributions of measurements. Finally, brief conclusions are presented in Sect. \ref{concl}.

\section{\label{extbek}Extended Bekenstein-type models}

In Bekenstein's class of varying $\alpha$ models \cite{Bekenstein}, a massless scalar field has a linear coupling to the $F^2$ term of the $U(1)$ gauge field, which we denote $\phi$; thus a change in the background value of $\phi$ leads to change of the effective value of $\alpha$ \cite{Carroll,Dvali,Chiba}. Importantly, Bekenstein also noticed that the $F^2$ has a non-vanishing matrix element over protons and neutrons, implying that the cosmological evolution of the field is driven by the baryon energy density.

Specifically, the electromagnetic part of the Lagrangian has the form
\be
{\cal L}= -\frac{1}{4}B_F(\phi)F_{\mu\nu}F^{\mu\nu}
\ee
where the gauge kinetic function $B_F$ can be linearized to
\be\label{gkf}
B_F(\phi)=1-\kappa\zeta_F(\phi-\phi_0)
\ee
with $\kappa^2=8\pi G$ and $\phi_0$ being the field value today. This linearization is expected to be a good approximation both because the field is expected to be moving slowly and lead to small variations of $\alpha$ (larger ones being experimentally and observationally ruled out, as we see in the next section) and because, as was pointed out in \cite{Dvali}, the absence of such a term would require the presence a $\phi\to-\phi$ symmetry, but such a symmetry must be broken throughout most of the cosmological evolution. It follows that the relative variation of $\alpha$, which is the observational parameter of choice, has the value
\be
\frac{\Delta\alpha}{\alpha}(z)\equiv\frac{\alpha(z)-\alpha_0}{\alpha_0}= \zeta_F\kappa(\phi-\phi_0)-1\,; \label{defalpha}
\ee
thus a negative value corresponds to a smaller value of $\alpha$ in the past (but note that this definition is opposite to the one used by \cite{OlivePospelov}). As is physically clear, the relevant parameter in the cosmological evolution is the field displacement relative to its present-day value, so without loss of generality we henceforth set $\phi_0=0$.

In these models the proton and neutron masses are also expected to vary, due to the electromagnetic corrections of their masses, and one relevant consequence of this fact is that local tests of the Equivalence Principle lead to the conservative constraint on the dimensionless electromagnetic coupling parameter (see \cite{Uzan} for an overview)
\begin{equation}
|\zeta_F|<10^{-3}\,;\label{localzeta}
\end{equation}
stronger constraints can be obtained at the cost of some model-dependence.

Olive and Pospelov \cite{OlivePospelov} have discussed extensions of the Bekenstein model by allowing couplings of the scalar field to both a dark matter candidate and to a dark energy one; specifically they take the latter to be a cosmological constant (an assumption that we retain in the present work). Denoting these respectively by $\eta_m$ and $\eta_\Lambda$, one obtains the following field equation
\be
\ddot \phi + 3H \dot \phi =-3H_0^2\left[\eta_m\Omega_m\left({a_0\over a}\right)^3 +\eta_\Lambda \Omega_\Lambda\right]\,,\label{fieldeqt}
\ee
where the $\Omega_i=\rho_i/\rho_{crit}$ are the present-day matter and cosmological densities, expressed as fractions of the critical density. As a caveat, we also note that there are in principle additional source terms driving the evolution of the scalar field, for example those proportional to $B_F'$ and $B_\Lambda'$, but the contributions of these terms are expected to be subdominant, at least in the low-redshift regime that we consider in the rest of the article.

Olive and Pospelov also discuss various model scenarios to which this phenomenological description is expected to apply. In addition to the original Bekenstein model and a supersymmetrized version thereof, examples include string-dilaton, Brans-Dicke and gaugino driven modulus models. We refer the reader to \cite{OlivePospelov} for detailed discussions of these scenarios. However, in what follows we limit ourselves to this phenomenological context, taking $\eta_m$ and $\eta_\Lambda$ to be nominally free parameters (irrespective of theoretical priors that they may have in specific models), to be constrained by current or forthcoming data. This is the purpose of the following two sections.

For our present purposes it is also convenient to express the above field equation as a function of redshift. Using the standard change of variables $d/dt=-H(1+z)d/dz$ one easily finds
\be
\phi'' + \left(\frac{d\ln{H}}{dz}-\frac{2}{1+z}\right)\phi'= -3\left[\frac{\eta_m\Omega_m(1+z)^3 +\eta_\Lambda}{(1+z)^2E^2(z)}\right]\,,\label{fieldeqz}
\ee
where
\be
E^2(z)=\frac{H^2(z)}{H_0^2}=\Omega_m(1+z)^3+\Omega_\Lambda\,.\label{friedmann}
\ee
In the last equality we are assuming that the Universe only includes non-relativistic matter and a cosmological constant (the radiation being neglected since we are concerned with low-redshift observations), and we further assume a flat universe (fully in agreement with the latest cosmological data \cite{Planck}), so $\Omega_m+\Omega_\Lambda=1$. As in the case of the original Bekenstein model, it is easy to check that observational constraints imply that the energy density of the field $\phi$ must be very small, yielding a negligible contribution to the Friedmann equation. Note that in this flat $\Lambda$CDM case there is an exact solution to the Friedmann equation, 
\be
\left(\frac{a(t)}{a_0}\right)^3= \frac{\Omega_m}{\Omega_\Lambda}\left[\sinh\left(\frac{3}{2}\Omega_\Lambda^{1/2}H_0t\right)\right]^2\,.
\ee
Interestingly, in this case the field equation can also be analytically integrated, and using Eq. (\ref{defalpha}) we obtain
\begin{widetext}
\be\label{maineq}
\frac{\Delta\alpha}{\alpha}(z)=2\zeta_m\log{(1+z)}+\frac{2(\zeta_\Lambda-2\zeta_m)}{3\sqrt{\Omega_\Lambda}}\left[\log{\left(\frac{1+\sqrt{\Omega_\Lambda}}{\sqrt{\Omega_m}}\right)}-\sqrt{E^2(z)}\log{\left(\frac{\sqrt{\Omega_\Lambda}+\sqrt{E^2(z)}}{\sqrt{\Omega_m(1+z)^3}}\right)}\right]\,;
\ee
\end{widetext}
here $\log$ denotes the natural logarithm, and we have defined
\be\label{defzetam}
\zeta_m\equiv\zeta_F\eta_m
\ee
\be\label{defzetal}
\zeta_\Lambda\equiv\zeta_F\eta_\Lambda
\ee
which are the parameter combinations that are constrained by the astrophysical measurements of $\alpha$. It is also interesting to consider the behavior of this solution in the two asymptotic limits. In the low redshift limit $z<<1$ we obtain a linearized behavior
\be
\frac{\Delta\alpha}{\alpha}(z)=\left[\zeta_\Lambda+(2\zeta_m-\zeta_\Lambda)\frac{\Omega_m}{\sqrt{\Omega_\Lambda}}\log{\frac{1+\sqrt{\Omega_\Lambda}}{\sqrt{\Omega_m}}} \right]z\,,
\ee
while at high redshifts ($z>>1$, but still neglecting the radiation density contribution) we obtain a dilaton-like logarithmic behavior
\be
\frac{\Delta\alpha}{\alpha}(z)=2\zeta_m \log{z}+\frac{2}{3}(\zeta_\Lambda-2\zeta_m)\left[\frac{1}{\sqrt{\Omega_\Lambda}}\log{\frac{1+\sqrt{\Omega_\Lambda}}{\sqrt{\Omega_m}}}-1\right]\,.
\ee
Also relevant is the present-day drift rate,
\be
\left(\frac{1}{H}\frac{\dot\alpha}{\alpha}\right)_0=-\zeta_\Lambda+2(\zeta_\Lambda-2\zeta_m)\frac{\Omega_m}{\sqrt{\Omega_\Lambda}}\log{\frac{1+\sqrt{\Omega_\Lambda}}{\sqrt{\Omega_m}}}\,,
\ee
which is constrained by laboratory tests using atomic clocks. As a simple illustration, if we take $\Omega_m=0.3$ this becomes
\be\label{sensclocks}
\left(\frac{1}{H}\frac{\dot\alpha}{\alpha}\right)_0=-0.13\zeta_\Lambda-1.74\zeta_m\,;
\ee
note that the sensitivity to the coupling $\zeta_\Lambda$ is much weaker than the sensitivity to $\zeta_m$. For comparison, if we write Eq. (\ref{maineq}) at redshifts $z=0.14$ and $z=1.15$, still assuming $\Omega_m=0.3$, we find
\be\label{sensoklo}
\left(\frac{\Delta\alpha}{\alpha}\right)_{z=0.14}=0.07\zeta_\Lambda+0.12\zeta_m\,
\ee
\be
\left(\frac{\Delta\alpha}{\alpha}\right)_{z=1.5}=0.27\zeta_\Lambda+1.29\zeta_m\,.
\ee
The former is the effective redshift of the Oklo bound (to be introduced in the next section), and the relatively low sensitivity to both couplings is worthy of note. This sensitivity increases with redshift, and by $z=1.5$, a value that is typical of spectroscopic measurements, these sensitivities are already comparable to those of atomic clocks. We therefore expect to obtain stronger constraints on $\zeta_m$ than on $\zeta_\Lambda$.

\begin{figure}
\includegraphics[width=0.5\textwidth]{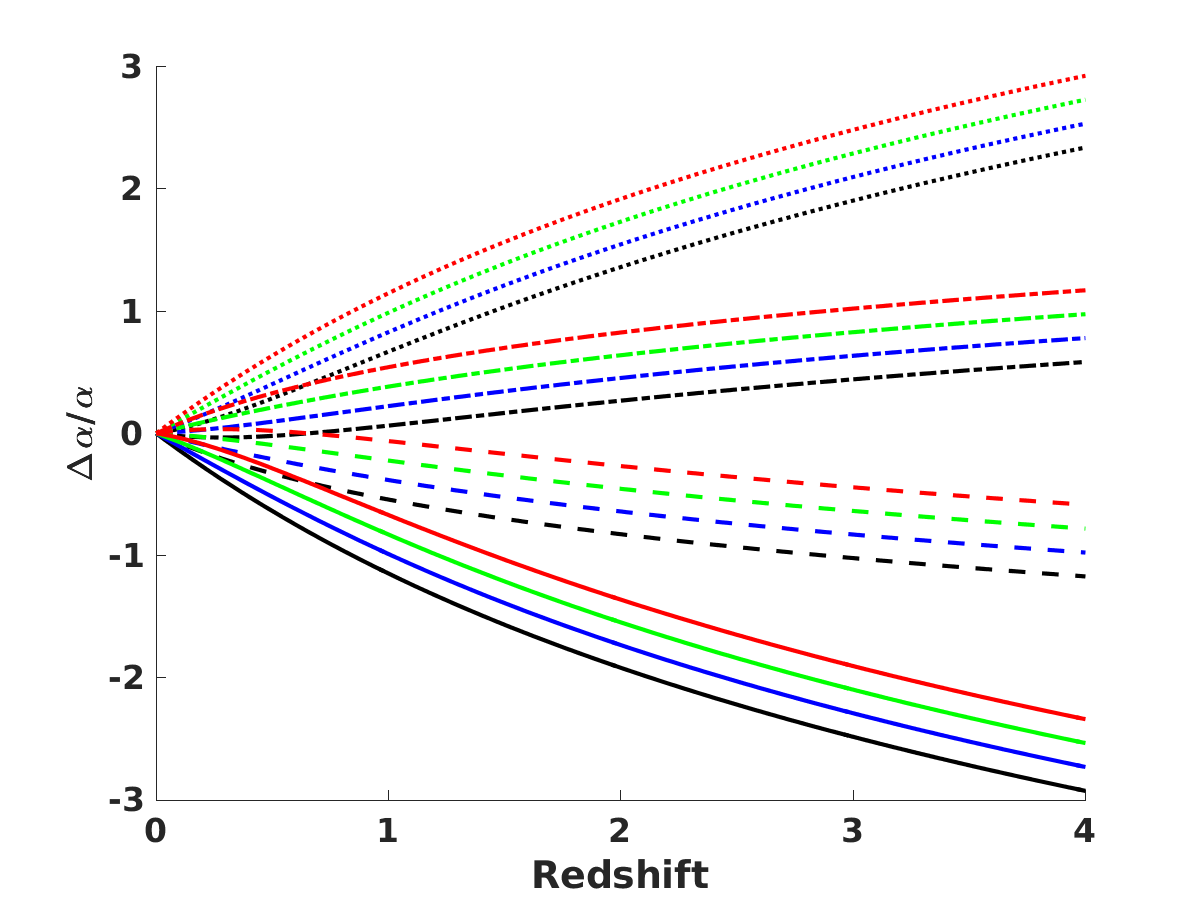}
\includegraphics[width=0.5\textwidth]{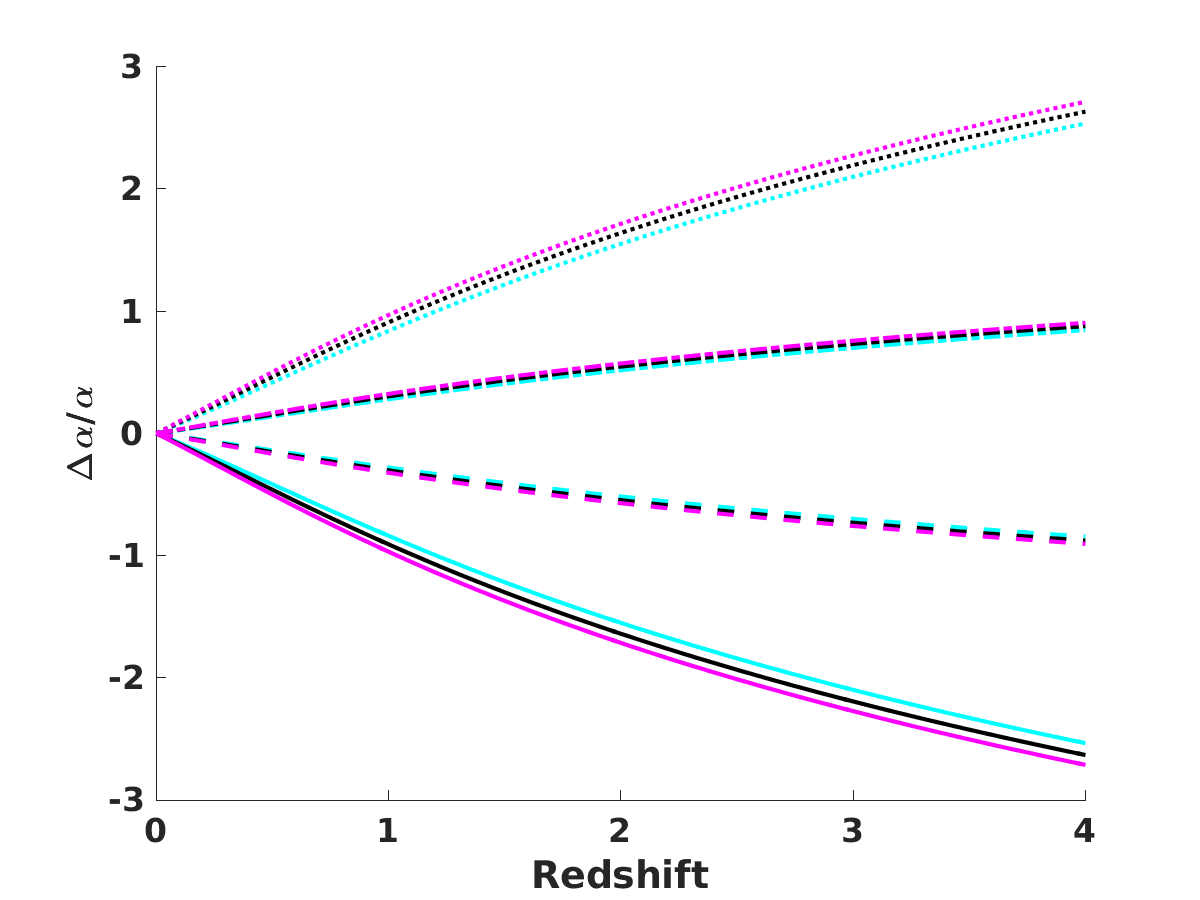}
\caption{Fine-structure constant evolution in the Olive-Pospelov model, if one assumes values of $\zeta_m$ of -1 (solid lines), -1/3 (dashed), +1/3 (dash-dotted) and +1 (dotted). {\bf Top panel:} Evolution for values of $\zeta_\Lambda$ of -1 (black lines), -1/3 (blue), +1/3 (green) and +1 (red); $\Omega_m=0.3$ was used throughout. {\bf Bottom panel:} Evolution for values of $\Omega_m$ of 0.25 (cyan lines), 0.30 (black), and 0.35 (magenta); $\zeta_\Lambda=0$ was used throughout.}
\label{fig1}
\end{figure}

Figure \ref{fig1} illustrates the redshift dependence of $\alpha$ in these models, and its sensitivity to the couplings $\zeta_m$ and $\zeta_\Lambda$ and the matter density $\Omega_m$; note the comparatively low sensitivity to the latter. For what follows it is also interesting to note that $\Delta\alpha/\alpha$ depends linearly on the couplings, and moreover their values determine the order of magnitude of the variation. Thus we expect that constraints on the couplings should be of the order of parts per million (ppm), which is the sensitivity of current astrophysical measurements of $\alpha$, as we discuss in the next section. For this reason we will express the constraints on the couplings in ppm units.

\section{\label{currdata}Current constraints}

We use all the available direct astrophysical measurements of $\alpha$ to constrain these models. We separately consider the Webb {\it et al.} \cite{Webb} data (a large data set of 293 archival data measurements obtained from the HIRES and UVES spectrographs) and the smaller but more recent data set of 21 dedicated measurements listed in Table \ref{tab1}. Further details on this compilation can be found in \cite{ROPP}.  Additionally we use the constraint from the Oklo natural nuclear reactor \cite{Oklo},
\begin{equation} \label{okloalpha}
\frac{\Delta\alpha}{\alpha} =0.005\pm0.061\, ppm\,,
\end{equation}
at an effective redshift $z=0.14$, and the atomic clocks laboratory bound on the current drift of $\alpha$ by Rosenband {\it et al.} \cite{Rosenband},
\be\label{rosen}
\left(\frac{1}{H}\frac{\dot\alpha}{\alpha}\right)_0=-0.22\pm0.32\, ppm\,;
\ee
recall that for convenience we are using units of parts per million throughout. Note that while the Oklo bound is nominally stronger, the atomic clocks bound is much more sensitive to the model couplings---compare Eqs. (\ref{sensclocks}) and (\ref{sensoklo}). This is important for what follows.

\begin{table}
\begin{center}
\begin{tabular}{|c|c|c|c|c|}
\hline
Object & z & ${ \Delta\alpha}/{\alpha}$ (ppm) & Spectrograph & Ref. \\
\hline
J0026$-$2857 & 1.02 & $3.5\pm8.9$ & UVES & \protect\cite{MalecNew} \\
J0058$+$0041 & 1.07 & $-1.4\pm7.2$ & HIRES & \protect\cite{MalecNew} \\
3 sources & 1.08 & $4.3\pm3.4$ & HIRES & \protect\cite{Songaila} \\
HS1549$+$1919 & 1.14 & $-7.5\pm5.5$ & UVES/HIRES/HDS & \protect\cite{LP3} \\
HE0515$-$4414 & 1.15 & $-1.4\pm0.9$ & UVES & \protect\cite{Kotus} \\
J1237$+$0106 & 1.31 & $-4.5\pm8.7$ & HIRES & \protect\cite{MalecNew} \\
HS1549$+$1919 & 1.34 & $-0.7\pm6.6$ & UVES/HIRES/HDS & \protect\cite{LP3} \\
J0841$+$0312 & 1.34 & $3.0\pm4.0$ & HIRES & \protect\cite{MalecNew} \\
J0841$+$0312 & 1.34 & $5.7\pm4.7$ & UVES & \protect\cite{MalecNew} \\
J0108$-$0037 & 1.37 & $-8.4\pm7.3$ & UVES & \protect\cite{MalecNew} \\
HE0001$-$2340 & 1.58 & $-1.5\pm2.6$ & UVES & \protect\cite{alphaAgafonova}\\
J1029$+$1039 & 1.62 & $-1.7\pm10.1$ & HIRES & \protect\cite{MalecNew} \\
HE1104$-$1805 & 1.66 & $-4.7\pm5.3$ & HIRES & \protect\cite{Songaila} \\
HE2217$-$2818 & 1.69 & $1.3\pm2.6$ & UVES & \protect\cite{LP1}\\
HS1946$+$7658 & 1.74 & $-7.9\pm6.2$ & HIRES & \protect\cite{Songaila} \\
HS1549$+$1919 & 1.80 & $-6.4\pm7.2$ & UVES/HIRES/HDS & \protect\cite{LP3} \\
Q1103$-$2645 & 1.84 & $3.5\pm2.5$ &  UVES & \protect\cite{Bainbridge}\\
Q2206$-$1958 & 1.92 & $-4.6\pm6.4$ &  UVES & \protect\cite{MalecNew}\\
Q1755$+$57 & 1.97 & $4.7\pm4.7$ & HIRES & \protect\cite{MalecNew} \\
PHL957 & 2.31 & $-0.7\pm6.8$ & HIRES & \protect\cite{MalecNew} \\
PHL957 & 2.31 & $-0.2\pm12.9$ & UVES & \protect\cite{MalecNew} \\
\hline
\end{tabular}
\caption{\label{tab1}Available dedicated measurements of $\alpha$. Listed are, respectively, the object along each line of sight, the redshift of the measurement, the measurement itself (in parts per million), the spectrograph, and the original reference. The third measurement is the weighted average from 8 absorbers along the lines of sight of HE1104-1805A, HS1700+6416 and HS1946+7658, reported in \cite{Songaila} without the values for the individual systems.}
\end{center}
\end{table}

\begin{figure*}
\includegraphics[width=0.45\textwidth]{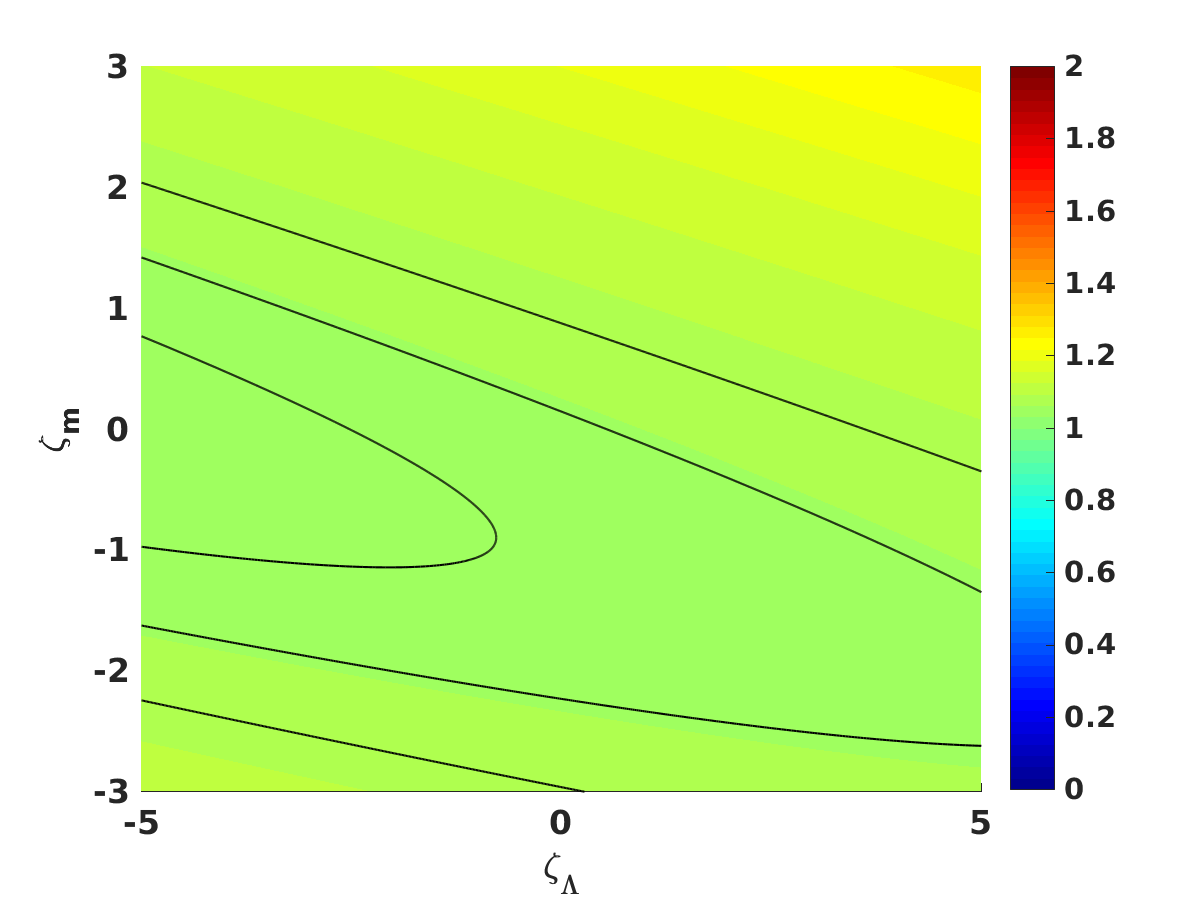}
\includegraphics[width=0.45\textwidth]{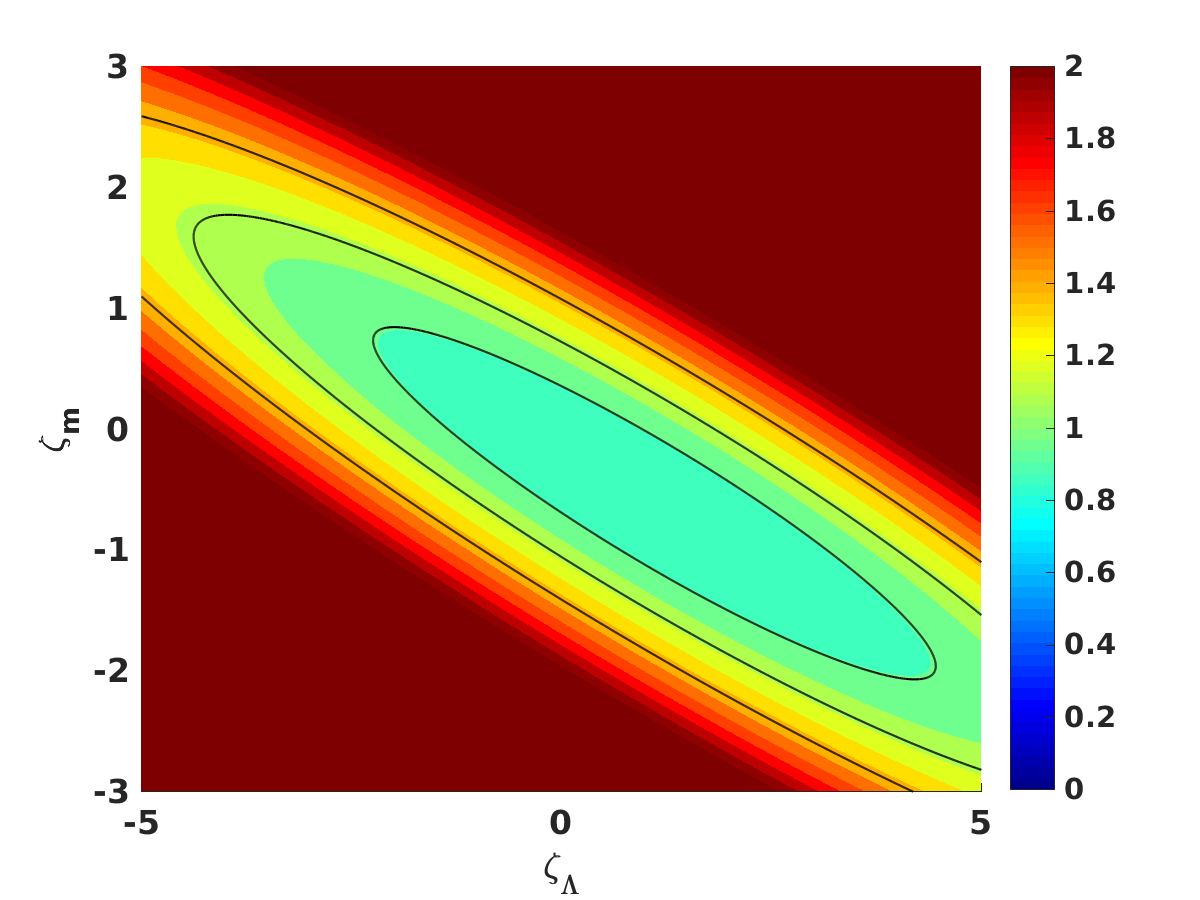}
\includegraphics[width=0.45\textwidth]{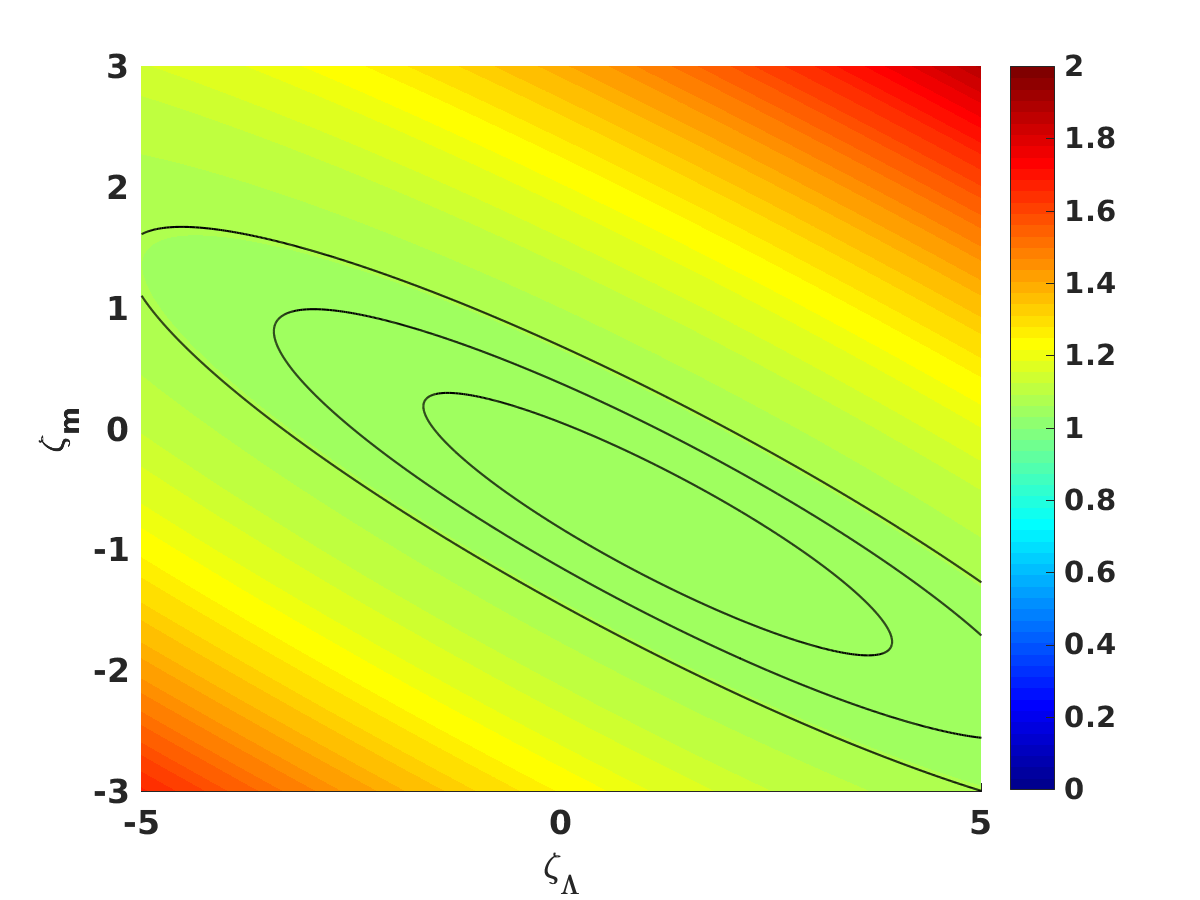}
\includegraphics[width=0.45\textwidth]{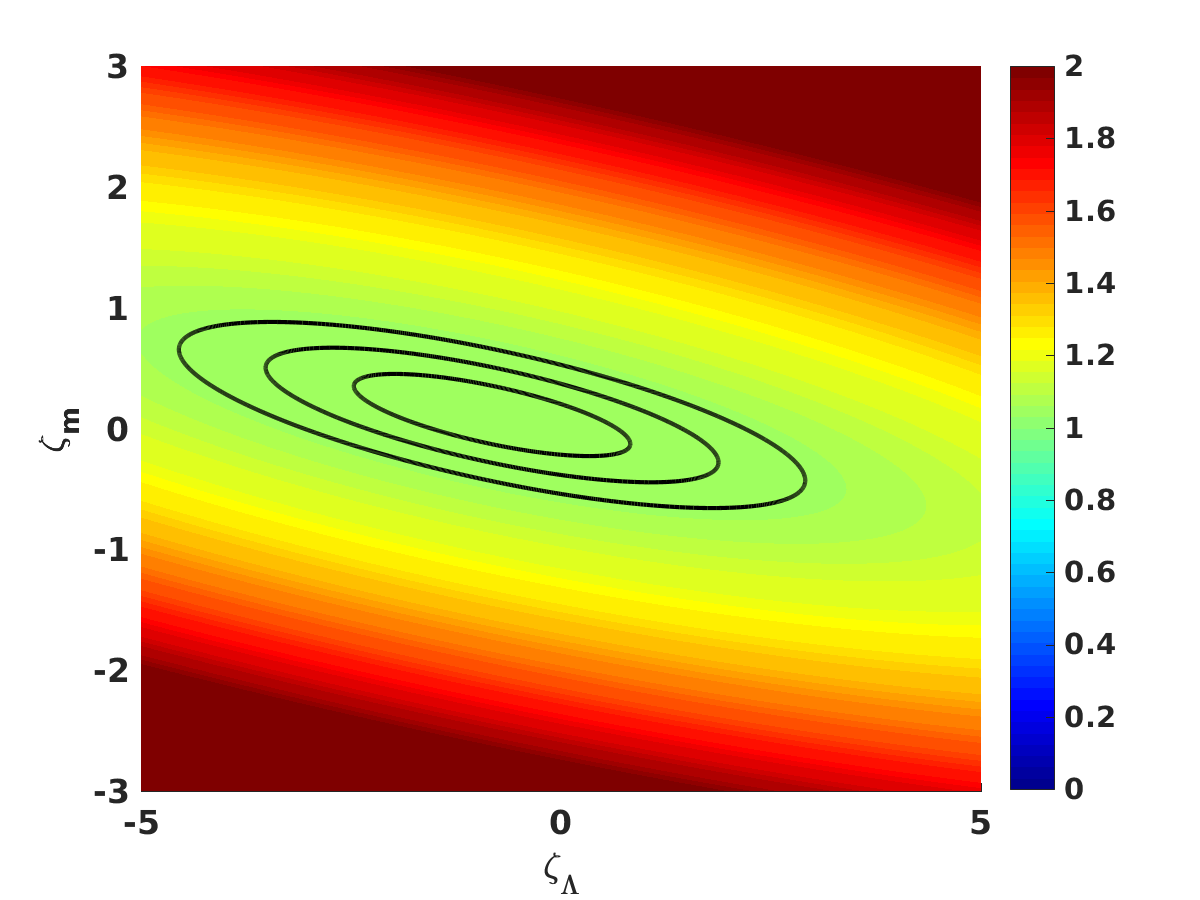}
\caption{Current constraints on Olive-Pospelov models, from the Webb {\it et al.} data (top left), the recent fine-structure constant measurements plus Oklo (top right), all the $z>0$ measurements combined (bottom left) and the complete data set including the atomic clock measurement (bottom right). In all cases the black lines correspond to the one, two and three sigma confidence levels, while the color map depicts the reduced chi-square. Moreover, to facilitate a visual comparison, the depicted range of each coupling is the same in all panels.}
\label{fig2}
\end{figure*}

\begin{figure}
\includegraphics[width=0.5\textwidth]{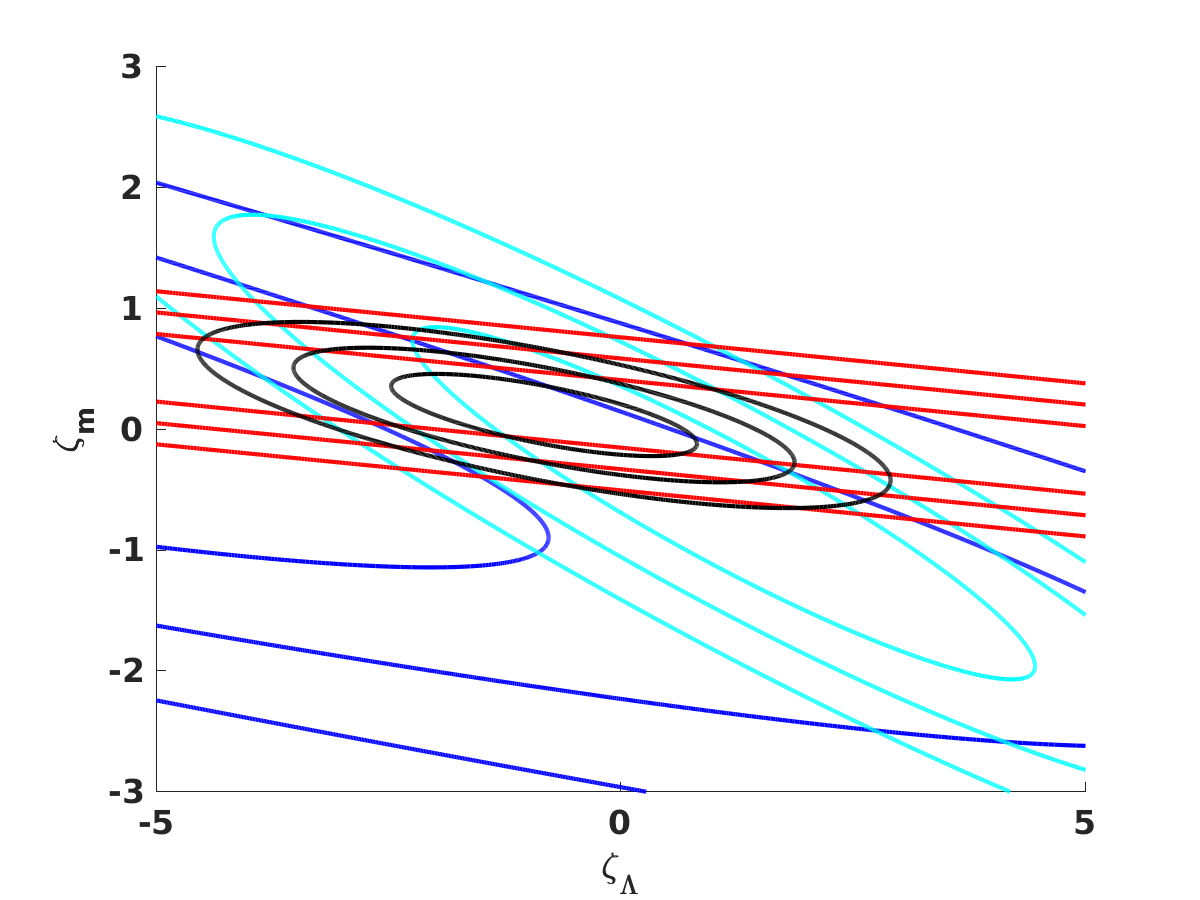}
\includegraphics[width=0.5\textwidth]{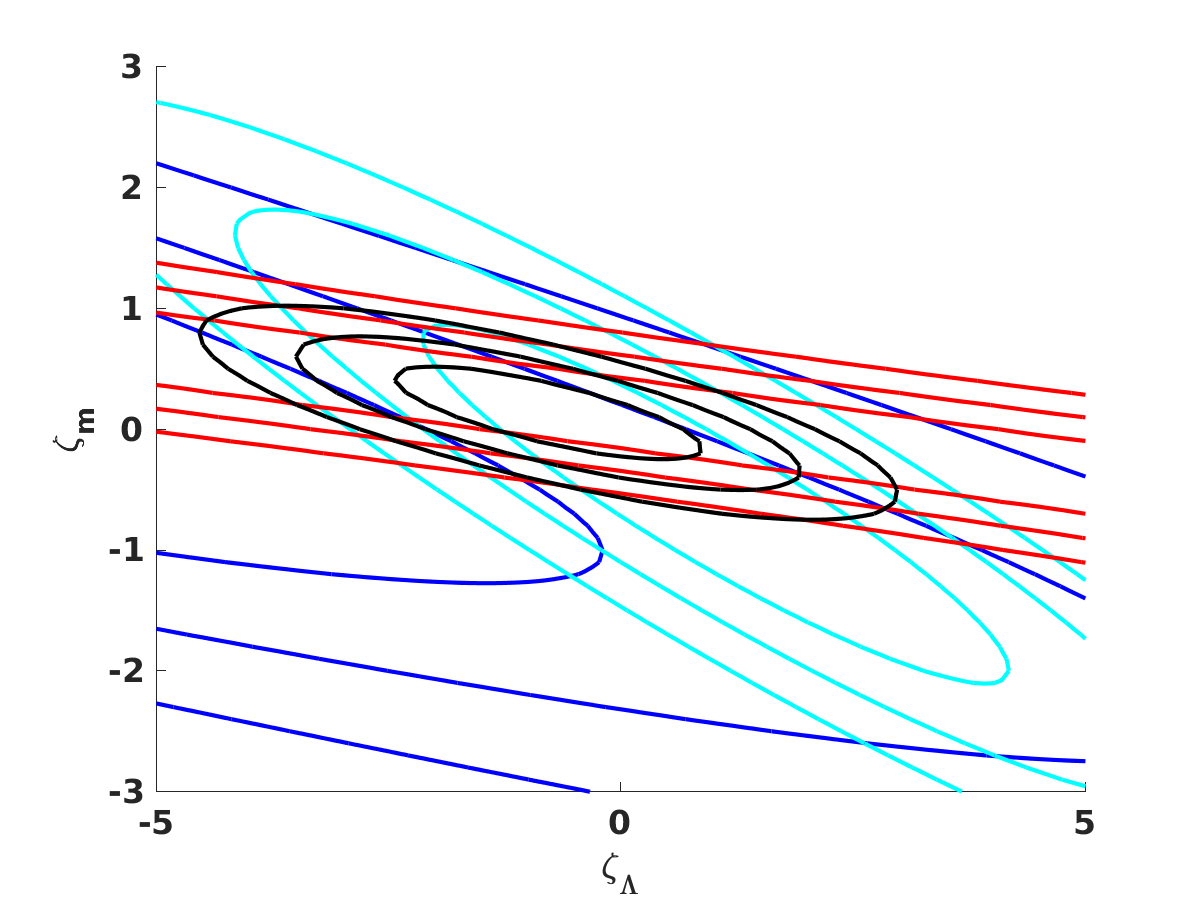}
\caption{Current constraints in the $\zeta_m$--$\zeta_\Lambda$ plane for the Olive-Pospelov models, from the Webb {\it et al.} data (blue lines), the recent $z>0$ fine-structure constant measurements (cyan), the atomic clock bound (red) and the complete data set (black). The one, two and three sigma confidence levels are shown in all cases. {\bf Top panel:} Matter density assumed to be perfectly known, with $\Omega_m=0.3$. {\bf Bottom panel:} Matter density allowed to vary and marginalized; the cosmological data sets described in the main text were also included in the analysis.}
\label{fig3}
\end{figure}

We start by making the simplifying assumption of fixed (perfectly known) matter density, specifically we choose $\Omega_m=0.3$ in agreement with the latest cosmological data \cite{Planck,Jones}. A standard likelihood analysis leads to the constraints in the $\zeta_m$--$\zeta_\Lambda$ plane shown in Fig.~ \ref{fig2}. As was found for the case of the standard Bekenstein model \cite{Leal,LeiteBek}, the Webb {\it et al.} data set has a mild (less than two standard deviations) statistical preference for non-zero couplings, but the other sub-sets and the combined data set are compatible with the null result. This is further illustrated in Fig. \ref{fig3}, which also confirms that the atomic clock bound mostly constrains $\zeta_m$ while being comparatively insensitive to $\zeta_\Lambda$.

We can nevertheless ask whether these results will be significantly changed if one allows the matter density to be a free parameter (while still assuming a flat universe). The expectation that this will not be the case stems from the fact that the correlation between the two couplings and the matter density is small, but nevertheless we have carried out this additional analysis. In this case, in addition to the aforementioned $\alpha$ measurements we use two background cosmology data sets: the Union2.1 set of 580 Type Ia supernovas \cite{Union} and a compilation of 38 Hubble parameter measurements by Farooq {\it et al.} \cite{Farooq}. Clearly the cosmological data sets mostly fix the matter density, while the $\alpha$ measurements constrain the two couplings.

\begin{figure}
\includegraphics[width=0.5\textwidth]{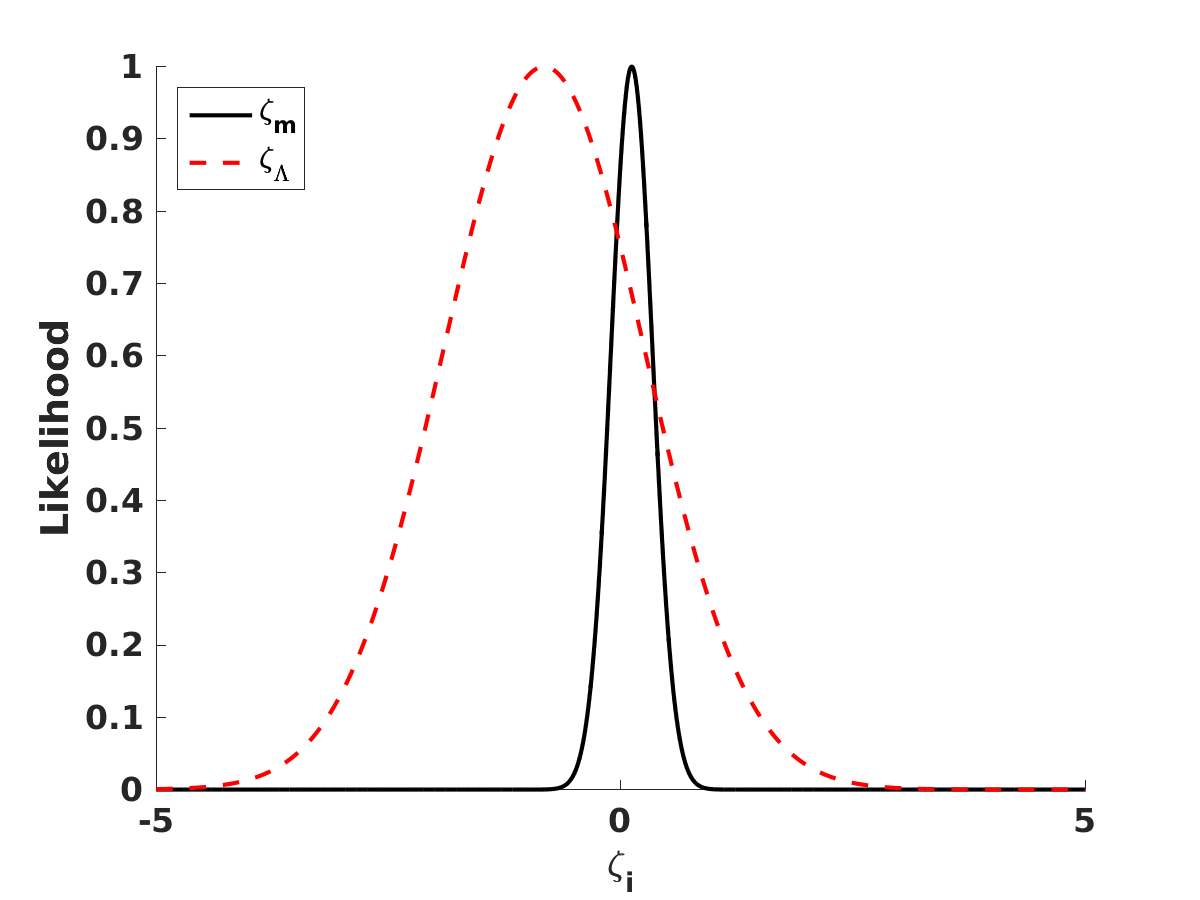}
\caption{One-dimensional likelihood for each of the couplings, with the other marginalized.}
\label{fig4}
\end{figure}

The two panels of Fig.~\ref{fig3} compare the constraints on the $\zeta_m$--$\zeta_\Lambda$ plane obtained either by keeping $\Omega_m$ fixed (as was already done for Fig.~\ref{fig2}) or by allowing it to vary and then marginalizing it, confirming that the differences are quite small. This is also the case for the one-dimensional marginalized likelihoods for each of the couplings, which are shown in Fig.~\ref{fig4}.


Table \ref{tab2} summarizes the current constraints. In addition to comparing the results obtained by fixing or marginalizing the matter density, the table also compares, for the former case, the constraints obtained with or without the atomic clocks bound. As is clear by comparing the bottom panels of Fig. \ref{fig2}, this does have a significant impact on the constraints: not only does the overall best fit change (by slightly more than one standard deviation), but the overall uncertainties are significantly reduced, especially in the case of $\zeta_m$. We thus confirm that current data can constrain both phenomenological couplings (each of which is the product of the electromagnetic coupling $\zeta_F$ and one of the model couplings $\eta_i$) to parts per million level, with $\zeta_m$ being more tightly constrained than $\zeta_\Lambda$ by a factor of about four.

\begin{table}
\centering
\begin{tabular}{|c|c|c|c|}
\hline
Cosmological model & $\Omega_m$ fixed & $\Omega_m$ fixed & $\Omega_m$ free \\
Atomic clocks included & No & Yes & Yes \\
\hline
$\chi^2_\nu$ & $1.03$ & $1.04$ & $0.96$ \\
\hline
$\zeta_m$ ($68.3\%$ c.l.) & $-0.78\pm0.72$ & $0.12\pm0.22$ & $0.1\pm0.3$ \\
$\zeta_m$ ($99.7\%$ c.l.) & $-0.78\pm2.14$ & $0.12\pm0.67$ & $0.1\pm0.8$ \\
\hline
$\zeta_\Lambda$ ($68.3\%$ c.l.) & $1.15\pm1.84$ & $-0.83\pm1.09$ & $-0.8\pm1.1$ \\
$\zeta_\Lambda$ ($99.7\%$ c.l.) & $1.15\pm5.53$ & $-0.83\pm3.26$ & $-0.8\pm3.3$ \\
\hline
\end{tabular}
\caption{\label{tab2}Current one and three sigma uncertainties on the couplings $\zeta_m$ and $\zeta_\Lambda$, marginalizing over the other parameter(s), obtained for various assumptions on the matter density (fixing it or allowing it to vary) and on the usage of the local atomic clocks bound, as described in the main text. The reduced chi-square of the best-fit parameters is also shown. Recall that the coupling values are expressed in parts per million.}
\end{table}

\section{\label{next}Forecasts for ESPRESSO and other future spectrographs}

At a time when this field is undergoing fast and significant observational developments, it is worth discussing how these developments will impact the constraints obtained in the previous section. Specifically, we are at the dawn of a new generation of high-resolution ultra-stable optical spectrographs, which will enable much more sensitive tests of the stability of $\alpha$ and other fundamental couplings. The first of these, ESPRESSO \cite{ESPRESSO}, built for the combined Coud\'e focus of ESO's VLT, is being commissioned at the time of writing (specifically, the commissioning phase is foreseen to take place from November 2017 to May 2018). The possibility of combining light from the four VLT unit telescopes means that ESPRESSO can effectively receive light from a 16 meter telescope. Thus ESPRESSO will become the instrument of choice for tests of the stability of fundamental couplings until the era of the Extremely Large Telescopes, and particularly its flagship spectrograph, ELT-HIRES \cite{HIRES}.

The ESPRESSO Cosmology and Fundamental Physics working group has made a preliminary selection of the list of targets for $\alpha$ measurements during the consortium's Guaranteed Time Observations (GTO). The criteria underlying this selection and the resulting list of targets are described in detail in \cite{MSC,LMMCC}. This consists of 14 absorption systems, in the redshift range $1.35\le z\le 3.02$, and broadly speaking these are the known quasar absorption systems that lead to the tightest constraints that can be observed from the VLT location (at Cerro Paranal, Chile) and with the ESPRESSO wavelength coverage---which is narrower than those of the VLT-UVES and Keck-HIRES spectrographs.

For this list of ESPRESSO targets we have generated simulated measurements with the expected ESPRESSO sensitivities, assuming two different scenarios: one with a fiducial model with no $\alpha$ variation (thus $\zeta_m=\zeta_\Lambda=0$), and the other with a fiducial model that is marginally inconsistent with current constraints, specifically $\zeta_m=\zeta_\Lambda=1$ ppm. In each of these cases we consider two sub-scenarios, which we call `Baseline' and `Ideal', which have also been discussed in \cite{LMMCC}. In practical terms, these correspond to assuming uncertainties in individual $\alpha$ measurements of 0.6 and 0.2 ppm, respectively. These are meant to represent two estimates of ESPRESSO's actual performance and sensitivity for these measurements, with the former being conservative and the latter being somewhat more optimistic (for example, reaching the corresponding signal-to-noise may require additional telescope time on each target). The actual performance of the instrument will only be known once the commissioning activities have been completed, but one may expect it to be somewhere between the two.

We also used these 14 targets in the ESPRESSO target list for a second specific forecast, for ELT-HIRES, by naively extrapolating the gains from the increased telescope collecting area and other foreseen technical improvements---we refer the reader to \cite{HIRES} for the instrument's Top Level Requirements. In these cases the `Baseline' and `Ideal' scenarios correspond to sensitivities a factor of six better than the ESPRESSO ones. We note that this forecast assumes that the wavelength coverage of ELT-HIRES will be at least as wide as that of ESPRESSO, and therefore allows it to observe all the ESPRESSO targets. This, as well as other technical specifications of the instrument (which is currently in its Phase A of development) is still under active study and consideration, and for this reason this forecast is necessarily more uncertain.

\begin{figure*}
\includegraphics[width=0.45\textwidth]{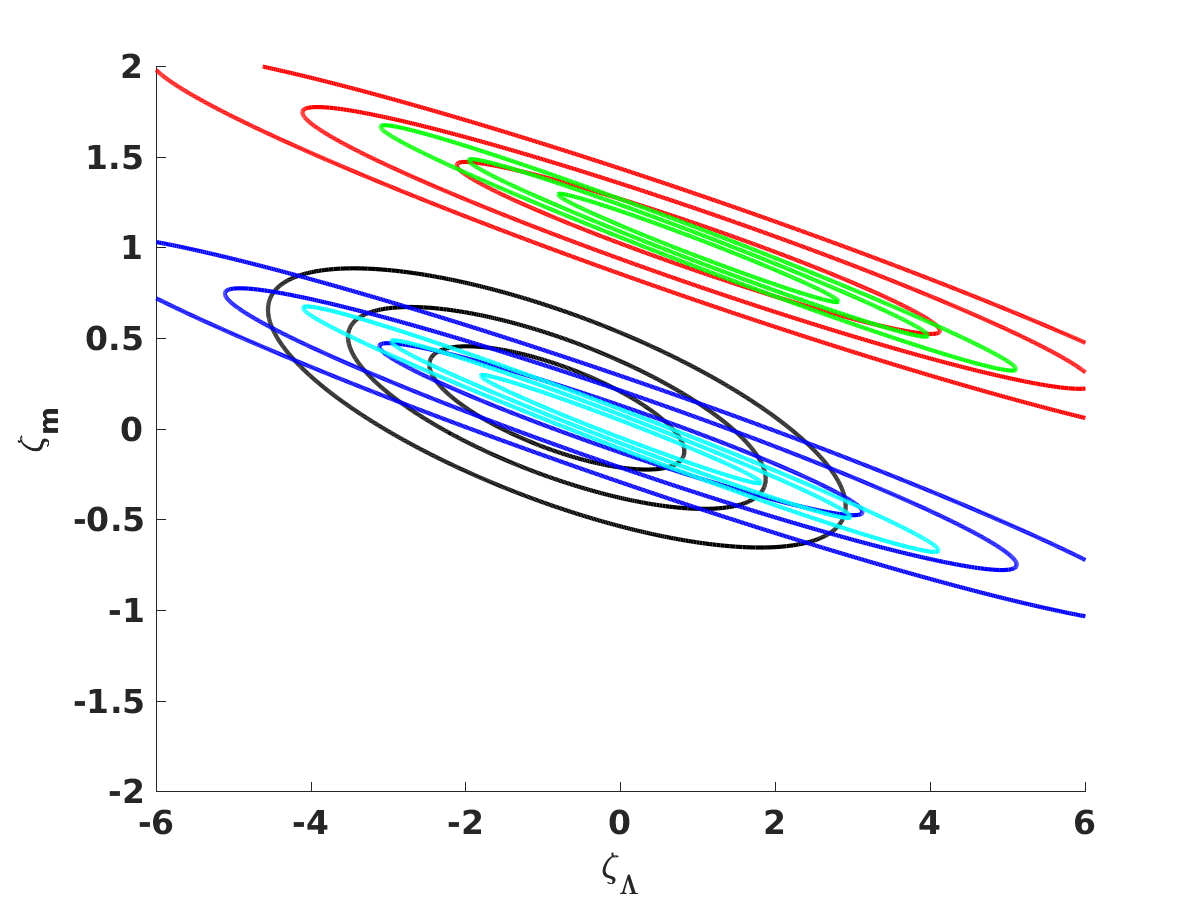}
\includegraphics[width=0.45\textwidth]{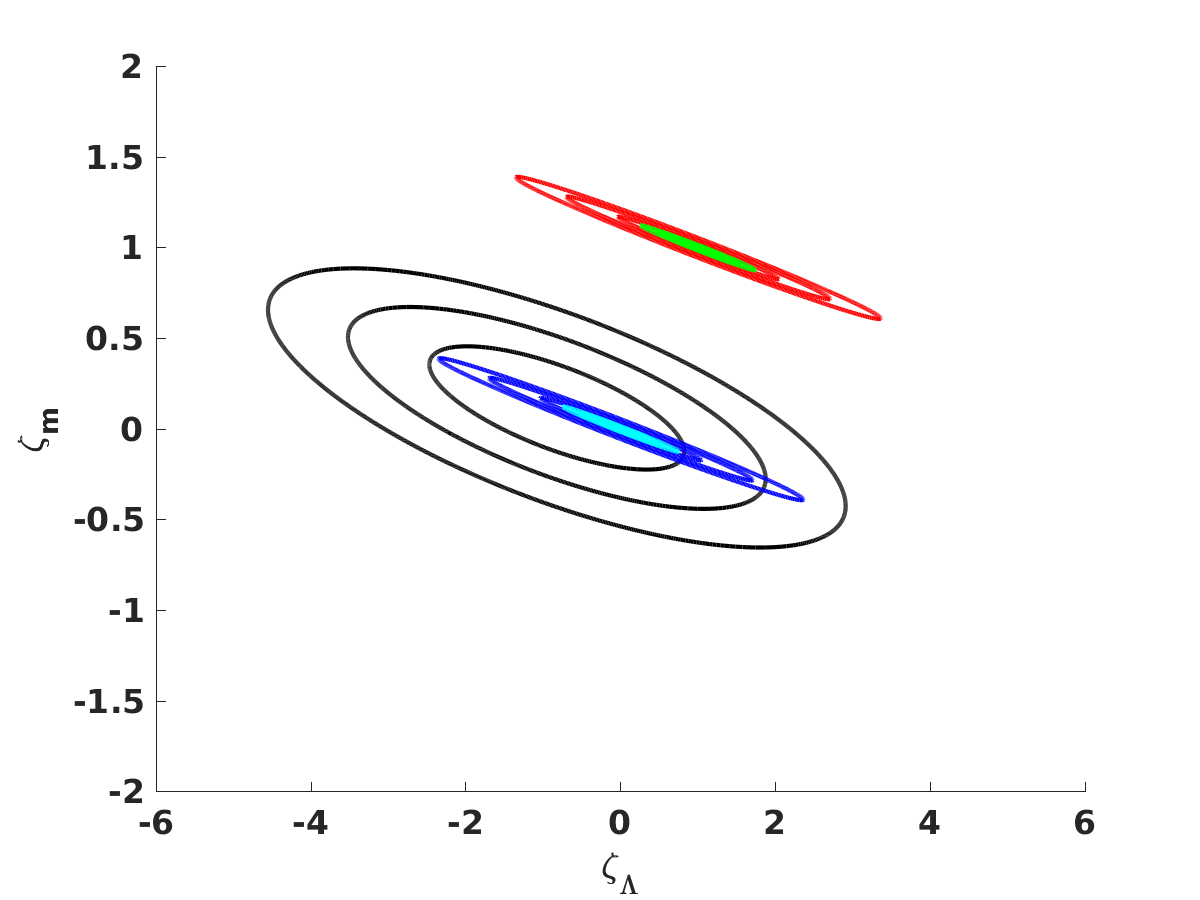}
\includegraphics[width=0.45\textwidth]{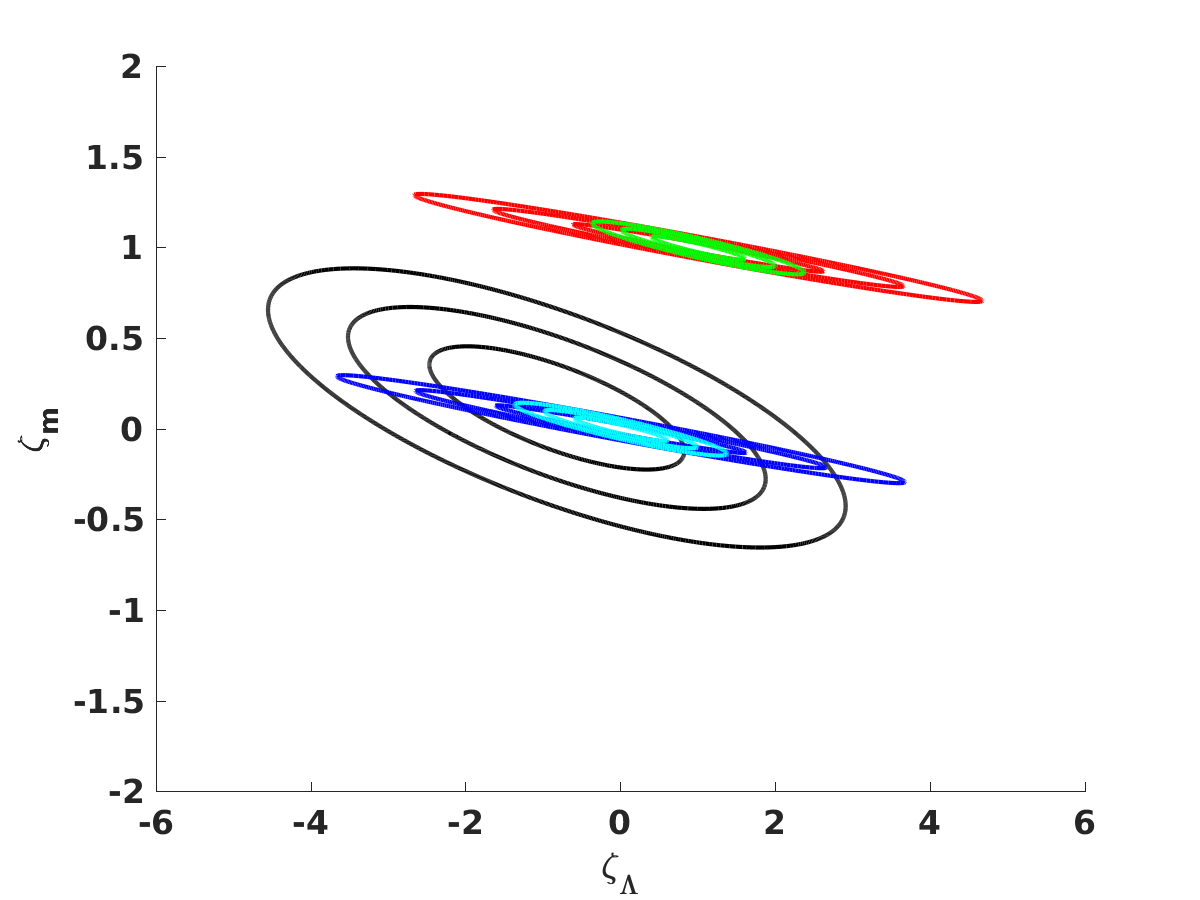}
\includegraphics[width=0.45\textwidth]{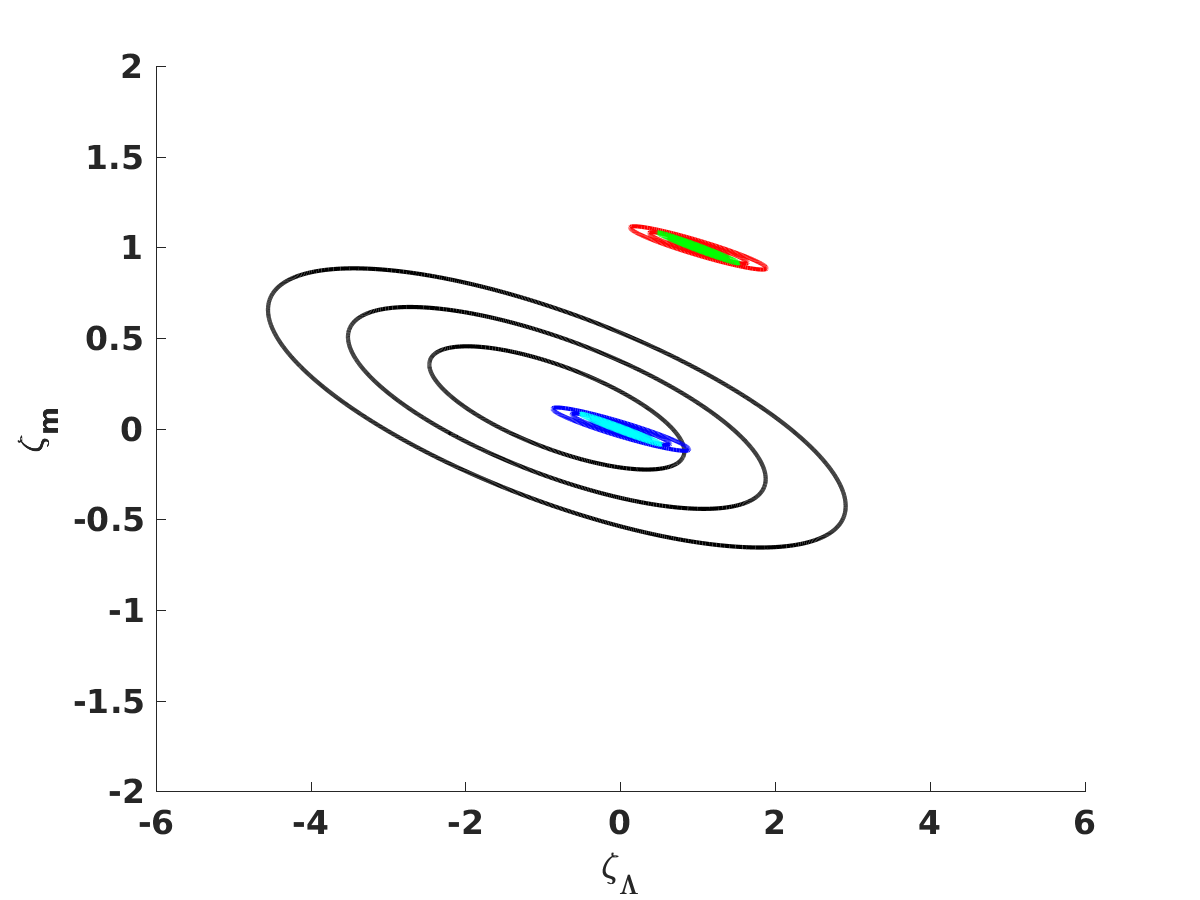}
\caption{Forecasted constraints on the Olive-Pospelov model, from the combination of the ESPRESSO fundamental physics GTO target list \protect\cite{MSC,LMMCC} with atomic clocks measurements. The top panels contain the forecasts including current atomic clock bounds, while the bottom panels contain forecasts with future (one order of magnitude better) atomic clock bounds. The left panels show the forecasts for ESPRESSO itself, while the right panels show forecasts for the same target list observed with ELT-HIRES. In all panels the black contours correspond to the current constraints obtained in Sect. \protect\ref{extbek}, the blue and cyan contours show the forecasts in the Baseline and Ideal scenarios for a fiducial model with $\zeta_m=\zeta_\Lambda=0$ while the red and green contours show the forecasts in the Baseline and Ideal scenarios for a fiducial model with $\zeta_m=\zeta_\Lambda=1$ ppm. One, two and three sigma contours are depicted throughout.}
\label{fig5}
\end{figure*}

In all cases we have generated a new (simulated) astrophysical data set, and used it instead of the current $\alpha$ measurements at non-zero redshift (that is, those from Webb {\it et al.}, Table \ref{tab1} and Oklo). Thus a data set of 315 current measurements is replaced by one with only 14, spanning a smaller redshift range---but naturally having much better precision. We have assumed a known matter density, $\Omega_m=0.3$, which was justified in the previous section. We have done the analysis both with and without the atomic clocks bound of Rosenband {\it et al.} \cite{Rosenband}, which as we previously saw is important in partially breaking the degeneracy between the two couplings. Finally, we have also studied an additional scenario, assuming that the current sensitivity of the atomic clock tests (cf. Eq. \ref{rosen}) is improved by one order of magnitude, leading to a sensitivity of $\sigma_{clocks}=0.032$ ppm.

\begin{table*}
\centering
\begin{tabular}{|l|c c|c c|c c|}
\hline
{ }  & \multicolumn{2}{|c|}{Without clocks} & \multicolumn{2}{|c|}{Current clocks} & \multicolumn{2}{|c|}{Future clocks}\\
Data set & $\delta\zeta_m$ & $\delta\zeta_\Lambda$ & $\delta\zeta_m$ & $\delta\zeta_\Lambda$ & $\delta\zeta_m$ & $\delta\zeta_\Lambda$ \\
\hline
Current constraints & $0.72$ & $1.84$ & $0.22$ & $1.09$ & N/A & N/A \\
\hline
ESPRESSO Baseline  & $0.73$ & $4.36$ & $0.31$ & $2.03$ & $0.08$ & $1.05$ \\
ESPRESSO Ideal  & $0.24$ & $1.45$ & $0.20$ & $1.19$ & $0.04$ & $0.39$ \\
\hline
ELT-HIRES Baseline  & $0.12$ & $0.73$ & $0.11$ & $0.68$ & $0.03$ & $0.25$ \\
ELT-HIRES Ideal  & $0.04$ & $0.24$ & $0.04$ & $0.24$ & $0.03$ & $0.16$ \\
\hline
\end{tabular}
\caption{\label{tab3}Current one sigma uncertainties on the couplings $\zeta_m$ and $\zeta_\Lambda$ (marginalizing the other) obtained in Sect. \protect\ref{currdata} from current data, and the corresponding forecasts for the ESPRESSO Fundamental Physics GTO target list and the forthcoming ELT-HIRES, under the assumptions discussed in the text. In particular, the cases with and without the current atomic clocks bound, as well as with an atomic clocks bound one order of magnitude more sensitive, are shown separately. As in the rest of the paper, uncertainties are in parts per million.}
\end{table*}

These forecasts are compared with the current constraints in Fig. \ref{fig5} (which for simplicity only shows the cases including the atomic clocks) and Table \ref{tab3} (which includes all the cases we have studied) . These make it clear that even a relatively small set of only 14 measurements will lead to very stringent constraints---and possibly even significant improvements. As previously mentioned the Baseline and Ideal scenarios are intended to bracket the actual performance of ESPRESSO and ELT-HIRES (with the caveat of larger uncertainties for the latter).

Broadly speaking we expect the ESPRESSO GTO data set by itself to lead to constraints at least as good as those of current astrophysical measurements, though at this level of sensitivity the atomic clocks bound of Rosenband {\it et al.} still leads to significant improvements. With the further gains in sensitivity expected for ELT-HIRES these would become strong enough to make the atomic clocks bound less important---on the assumption that the latter would not improve. However, this is clearly a pessimistic scenario. The atomic clocks bound is also expected to improve in the coming years, and we can see that our assumed improvement of one order of magnitude in the sensitivity of this bound again has a significant impact on the constraints, the reason being the previously discussed one: the high sensitivity of this bound to the coupling $\zeta_m$ partially breaks the degeneracy with $\zeta_\Lambda$.

\begin{figure}
\includegraphics[width=0.5\textwidth]{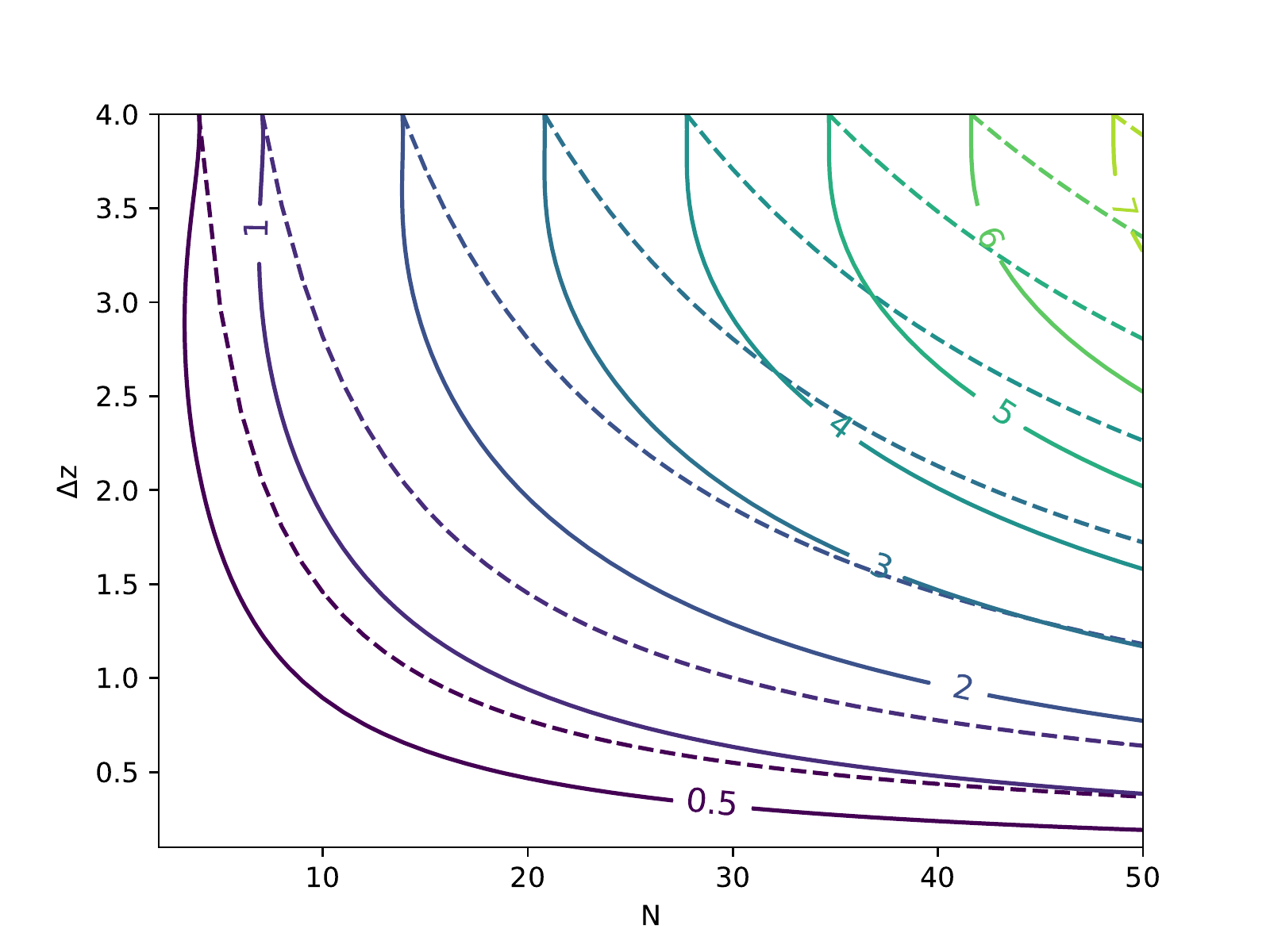}
\caption{Forecasted constraints on the Olive-Pospelov model for measurements of $\alpha$ with the ESPRESSO baseline sensitivity (that is, assuming a 0.6 ppm). The plot shows the value of the FoM (defined in the text) as a function of the number of measurements (shown on the horizontal axis), assumed to be uniformly distributed in a given redshift range (shown on the vertical axis). For the solid lines the redshift range is always centered at $z=2$, while for the dashed ones it always starts at $z=0$. For comparison, the FoM for the 14 ESPRESSO GTO targets is forecasted to be 1.03.}
\label{fig6}
\end{figure}
\begin{figure}
\includegraphics[width=0.5\textwidth]{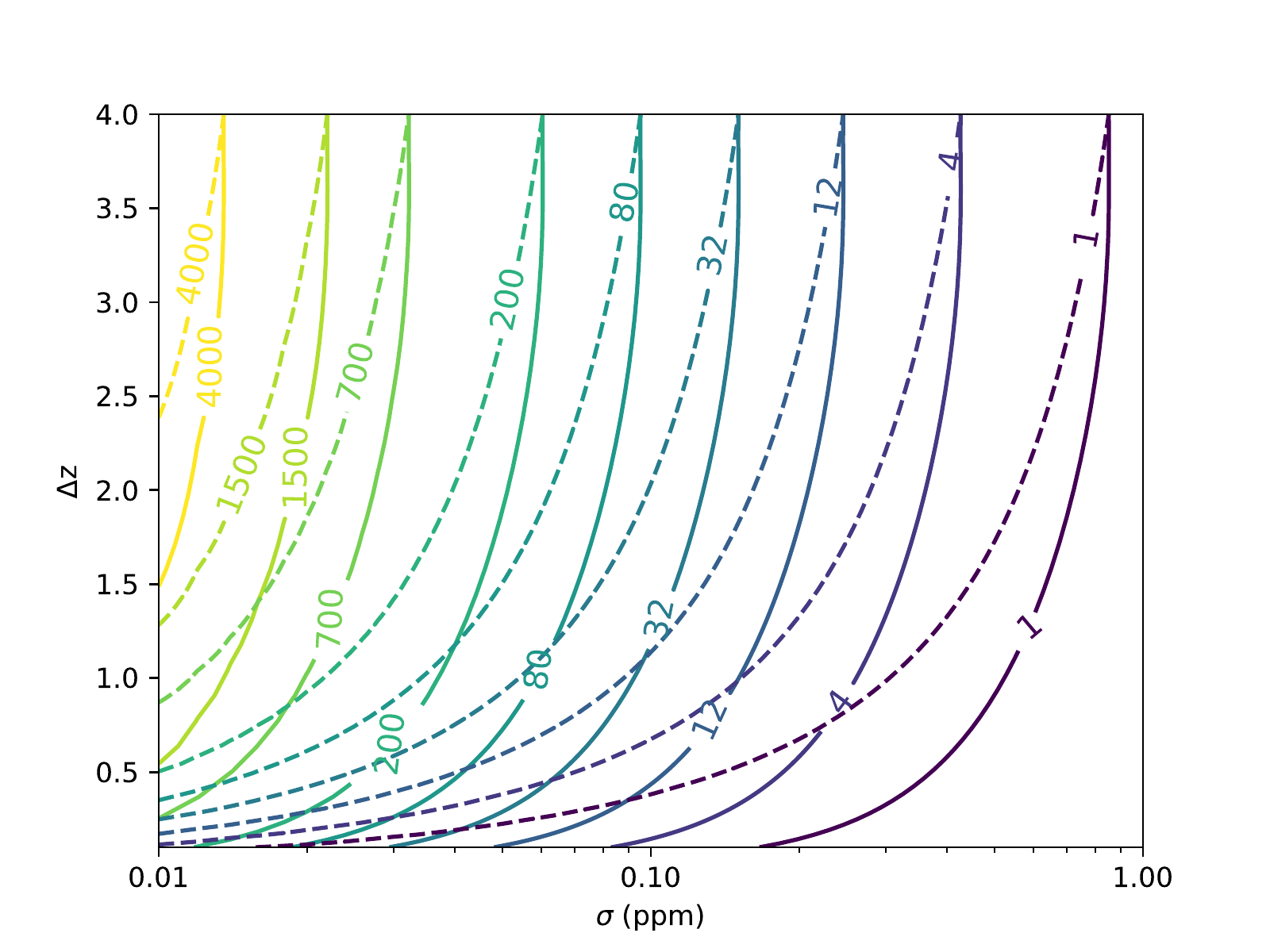}
\caption{Forecasted constraints on the Olive-Pospelov model from various sets of 14 astrophysical measurements of $\alpha$. The plot shows the value of the FoM (defined in the text) as a function of the uncertainty of each measurement (in ppm, shown on the horizontal axis), assumed to be uniformly distributed in a given redshift range (shown on the vertical axis). For the solid lines the redshift range is always centered at $z=2$, while for the dashed ones it always starts at $z=0$. For comparison, the FoM for the 14 ESPRESSO GTO targets is forecasted to be 1.03. }
\label{fig7}
\end{figure}

It is worthy of note that for this model Fisher Matrix based forecasts can be done in a particularly simple and generic way: since the $\alpha$ variation depends linearly on the couplings, the Fisher Matrix will not depend on the choice of the fiducial values of these couplings (for a fixed choice of cosmology, specifically of $\Omega_m$). One can therefore explore more general forecasts, discussing how constraints on the two couplings of this model depend on various observational parameters. The figure of merit (FoM) for this comparison is $\pi/A$, with $A$ being the area of the one-sigma confidence level ellipse in the $\zeta_m$--$\zeta_\Lambda$ plane. The two couplings are expressed in ppm units as before. In this case, since one of the parameters that we are varying is the number of measurements, we calculate the FoM using these measurements alone.

Figure \ref{fig6} shows the values of the FoM obtained by assuming a set of future measurements of $\alpha$ each of which has the ESPRESSO baseline sensitivity (that is, a 0.6 ppm uncertainty is assumed for each of the measurements). The FoM is shown as a function of the number of measurements and of the redshift range that they span. For simplicity we always assume that the measurements are uniformly distributed in the given redshift range. We further consider two alternative scenarios: a redshift range that is centered at $z=2$ (so for example a range $\Delta z=2$ corresponds to measurements in the range $1<z<3$), or a range that always starts at $z=0$ (so for example a range $\Delta z=2$ corresponds to measurements in the range $0<z<2$). The former, which is depicted by solid lines, is observationally motivated since $z=2$ is approximately the median redshift at which ESPRESSO can carry out these measurements \cite{MSC,LMMCC}. The latter, depicted by dashed lines, provides an interesting contrast and is also motivated by the fact that the observational cost of doing these measurements (in other words, the amount of telescope time needed to reach a given signal to noise) is smaller at lower redshifts, since lower redshift targets are typically brighter. Note that the two scenarios coincide for $\Delta z=4$: in that case both of them correspond to the redshift range $0<z<4$.

This analysis confirms the obvious point that larger numbers of measurements lead to larger FoMs (and hence stronger constraints), but they also show that a wide redshift coverage is important in this regard. For comparison, the FoM for the 14 targets of ESPRESSO GTO, which span the redshift range $1.35\le z\le 3.02$ (although they are not uniformly distributed in this redshift range) is forecasted to be 1.03, while the FoM for the current data is respectively 1.81 for all available data, and 0.33 if one excludes the atomic clocks bound from the analysis.

Importantly, the figure makes it clear that concentrating any number of measurements in a very narrow redshift range, or even taking the extreme scenario where all the available telescope time is spent doing the best possible measurements on a single target, which corresponds to taking $\Delta z\longrightarrow 0$ in the figure, always leads to a FoM below unity. Note this statement applies whether this is done at $z\sim0$ or at $z=2$, and clearly it applies to any redshift. Physically the reason is straightforward: in this class of models the behavior of $\alpha$ depends on two couplings, and measurements at a single redshift cannot disentangle them. On the other hand, as was pointed out in Sect. \ref{extbek}, the redshift dependence of $\alpha$ on the two couplings is itself redshift-dependent, and it is qualitatively different at low redshifts ($z<<1$, when the universe is accelerating) and at high redshifts (deep in the matter era). Having measurements that constrain both of these regimes is therefore an important consideration when trying to maximize the constraining power of the (effectively) limited amount of telescope time available for these measurements.

Finally, in Fig. \ref{fig7} we assume that the data set always has the same number of measurements as in the ESPRESSO GTO case (in other words, $N=14$), but allow their uncertainties (which are always the same within a given data set) to vary, ranging between 0.01 and 1 ppm. The latter sensitivity has already been achieved in one case for $\alpha$ measurements (refer to Table \ref{tab1}) and should be routinely achieved in the future for very large numbers of targets, while the former one can currently only be achieved within the Galaxy \cite{Levshakov} but is within the reach of next-generation facilities, at least for relatively small numbers of particularly suitable targets. As in the case of the previous figure, we assume that these measurements are uniformly distributed in the various redshift ranges, and consider both the cases of redshift ranges centered at $z=2$ and starting at $z=0$. The results of this analysis confirm that a wide redshift coverage is crucial regardless of the sensitivity of the individual measurements, and also make it clear that dedicated surveys with next-generation instruments, either doing high sensitivity (i.e., high signal to noise) measurements on selected numbers of targets or measuring significantly larger numbers of targets than the 14 of the ESPRESSO GTO---and in any case covering a wide redshift range---can improve by several orders of magnitude the FoM for the constraints on this class of models, corresponding to more than an order of magnitude improvement on the individual marginalized constraints on each of the couplings.

\section{\label{concl}Conclusions}

We have used currently available astrophysical tests of the stability of the fine-structure constant $\alpha$, together with the Oklo and local atomic clock constraints, to derive phenomenological constraints on an extension of the Bekenstein-type models for varying $\alpha$, first studied by Olive and Pospelov. We note that both the original model and its extensions can be seen as $\Lambda$CDM-like models with an additional dynamical degree of freedom, whose dynamics drives the evolution of $\alpha$ without having a significant impact on the Universe's dynamics. In particular, this degree of freedom cannot be responsible for the dark energy (for which a cosmological constant is still invoked). This is by no means a generic situation: there are many examples of dark energy models where the same dynamical degree of freedom can account both for the dark energy and for a varying $\alpha$---examples of such models (including current constraints on them) can be found in a recent review \cite{ROPP}. The question of which of these broad classes of models for varying $\alpha$ is the better motivated one is open to debate, but it is also beyond the scope of the present work. In any case, we emphasize that these are interesting models---at the very least at the phenomenological level at which we have considered them here---and are subject to stringent constraints. Moreover, should evidence for $\alpha$ variations be confirmed by future observations one can, in principle, distinguish between the two classes of models \cite{ROPP}.

Our analysis shows that the two model parameters $\zeta_m$ and $\zeta_\Lambda$ are currently constrained to ppm level. Recall from Sect.\ref{extbek} that these parameters are the product of the electromagnetic coupling $\zeta_F$, defined in Eq, \ref{gkf}, with the scalar field coupling to dark matter and dark energy, respectively denoted $\eta_m$ and $\eta_\Lambda$---cf. Eqs. \ref{defzetam} and \ref{defzetal}. Moreover, Local Equivalence principle tests require that $|\zeta_F|<10^{-3}$ (cf. Eq. \ref{localzeta}). Therefore, if one assumes that the value of $\zeta_F$ saturates this bound, then $\eta_m$ and $\eta_\Lambda$ are themselves constrained to be less that $10^{-3}$; however, the former assumption need not be true: if $\zeta_F$ is smaller, then the $\eta_i$ are allowed to be larger. It will be interesting to separately constrain all three parameters (as opposed to the two products of them) by combining local experiments and astrophysical observations. We leave this task for subsequent work.

In addition to providing current constraints on these models, we have also discussed how these are expected to improve in the context of the new generation of high-resolution ultra-stable spectrographs (of which ESPRESSO is the first example) and of expected improvements in local atomic clock tests. We have therefore used the current target list of 14 $\alpha$ measurements foreseen for the ESPRESSO GTO as a benchmark for these forecasts, while also considering more generic scenarios. We emphasize that an assumption of 14 targets is quite conservative: although they are, broadly speaking, the best targets available to ESPRESSO \cite{MSC,LMMCC}, further improvements can certainly come from observing additional targets. Indeed, about 300 different absorption systems have so far provided $\alpha$ measurements, and, in principle, the number of quasar absorption systems that can yield measurements is at least one thousand. Although these observations are costly in terms of telescope time, much larger data sets can therefore be put together in due course, leading to improvements on current constraints on each coupling by more than an order of magnitude.

\begin{acknowledgments}

This work was done in the context of project PTDC/FIS/111725/2009 (FCT, Portugal), with additional support from grant UID/FIS/04434/2013. Some of the work described herein was done at or following AstroCamp 2017.

CSA was partially supported by the Scientific Initiation grant CIAAUP-06/2017-BIC. ACL is supported by an FCT fellowship (SFRH/BD/113746/2015), under the FCT PD Program PhD::SPACE (PD/00040/2012). CJM is supported by an FCT Research Professorship, contract reference IF/00064/2012, funded by FCT/MCTES (Portugal) and POPH/FSE (EC).

\end{acknowledgments}

\bibliography{newbek}

\begin{thebibliography}{29}%
\makeatletter
\providecommand \@ifxundefined [1]{%
 \@ifx{#1\undefined}
}%
\providecommand \@ifnum [1]{%
 \ifnum #1\expandafter \@firstoftwo
 \else \expandafter \@secondoftwo
 \fi
}%
\providecommand \@ifx [1]{%
 \ifx #1\expandafter \@firstoftwo
 \else \expandafter \@secondoftwo
 \fi
}%
\providecommand \natexlab [1]{#1}%
\providecommand \enquote  [1]{``#1''}%
\providecommand \bibnamefont  [1]{#1}%
\providecommand \bibfnamefont [1]{#1}%
\providecommand \citenamefont [1]{#1}%
\providecommand \href@noop [0]{\@secondoftwo}%
\providecommand \href [0]{\begingroup \@sanitize@url \@href}%
\providecommand \@href[1]{\@@startlink{#1}\@@href}%
\providecommand \@@href[1]{\endgroup#1\@@endlink}%
\providecommand \@sanitize@url [0]{\catcode `\\12\catcode `\$12\catcode
  `\&12\catcode `\#12\catcode `\^12\catcode `\_12\catcode `\%12\relax}%
\providecommand \@@startlink[1]{}%
\providecommand \@@endlink[0]{}%
\providecommand \url  [0]{\begingroup\@sanitize@url \@url }%
\providecommand \@url [1]{\endgroup\@href {#1}{\urlprefix }}%
\providecommand \urlprefix  [0]{URL }%
\providecommand \Eprint [0]{\href }%
\providecommand \doibase [0]{http://dx.doi.org/}%
\providecommand \selectlanguage [0]{\@gobble}%
\providecommand \bibinfo  [0]{\@secondoftwo}%
\providecommand \bibfield  [0]{\@secondoftwo}%
\providecommand \translation [1]{[#1]}%
\providecommand \BibitemOpen [0]{}%
\providecommand \bibitemStop [0]{}%
\providecommand \bibitemNoStop [0]{.\EOS\space}%
\providecommand \EOS [0]{\spacefactor3000\relax}%
\providecommand \BibitemShut  [1]{\csname bibitem#1\endcsname}%
\let\auto@bib@innerbib\@empty
\bibitem [{\citenamefont {Webb}\ \emph {et~al.}(2011)\citenamefont {Webb},
  \citenamefont {King}, \citenamefont {Murphy}, \citenamefont {Flambaum},
  \citenamefont {Carswell},\ and\ \citenamefont {Bainbridge}}]{Webb}%
  \BibitemOpen
  \bibfield  {author} {\bibinfo {author} {\bibfnamefont {J.~K.}\ \bibnamefont
  {Webb}}, \bibinfo {author} {\bibfnamefont {J.~A.}\ \bibnamefont {King}},
  \bibinfo {author} {\bibfnamefont {M.~T.}\ \bibnamefont {Murphy}}, \bibinfo
  {author} {\bibfnamefont {V.~V.}\ \bibnamefont {Flambaum}}, \bibinfo {author}
  {\bibfnamefont {R.~F.}\ \bibnamefont {Carswell}}, \ and\ \bibinfo {author}
  {\bibfnamefont {M.~B.}\ \bibnamefont {Bainbridge}},\ }\href {\doibase
  10.1103/PhysRevLett.107.191101} {\bibfield  {journal} {\bibinfo  {journal}
  {Phys. Rev. Lett.}\ }\textbf {\bibinfo {volume} {107}},\ \bibinfo {pages}
  {191101} (\bibinfo {year} {2011})},\ \Eprint {http://arxiv.org/abs/1008.3907}
  {arXiv:1008.3907 [astro-ph.CO]} \BibitemShut {NoStop}%
\bibitem [{\citenamefont {Martins}(2017)}]{ROPP}%
  \BibitemOpen
  \bibfield  {author} {\bibinfo {author} {\bibfnamefont {C.~J. A.~P.}\
  \bibnamefont {Martins}},\ }\href {\doibase 10.1088/1361-6633/aa860e}
  {\bibfield  {journal} {\bibinfo  {journal} {Rept. Prog. Phys.}\ }\textbf
  {\bibinfo {volume} {80}},\ \bibinfo {pages} {126902} (\bibinfo {year}
  {2017})},\ \Eprint {http://arxiv.org/abs/1709.02923} {arXiv:1709.02923
  [astro-ph.CO]} \BibitemShut {NoStop}%
\bibitem [{\citenamefont {Bekenstein}(1982)}]{Bekenstein}%
  \BibitemOpen
  \bibfield  {author} {\bibinfo {author} {\bibfnamefont {J.~D.}\ \bibnamefont
  {Bekenstein}},\ }\href {\doibase 10.1103/PhysRevD.25.1527} {\bibfield
  {journal} {\bibinfo  {journal} {Phys. Rev.}\ }\textbf {\bibinfo {volume}
  {D25}},\ \bibinfo {pages} {1527} (\bibinfo {year} {1982})}\BibitemShut
  {NoStop}%
\bibitem [{\citenamefont {Sandvik}\ \emph {et~al.}(2002)\citenamefont
  {Sandvik}, \citenamefont {Barrow},\ and\ \citenamefont {Magueijo}}]{SBM}%
  \BibitemOpen
  \bibfield  {author} {\bibinfo {author} {\bibfnamefont {H.~B.}\ \bibnamefont
  {Sandvik}}, \bibinfo {author} {\bibfnamefont {J.~D.}\ \bibnamefont {Barrow}},
  \ and\ \bibinfo {author} {\bibfnamefont {J.}~\bibnamefont {Magueijo}},\
  }\href {\doibase 10.1103/PhysRevLett.88.031302} {\bibfield  {journal}
  {\bibinfo  {journal} {Phys. Rev. Lett.}\ }\textbf {\bibinfo {volume} {88}},\
  \bibinfo {pages} {031302} (\bibinfo {year} {2002})},\ \Eprint
  {http://arxiv.org/abs/astro-ph/0107512} {arXiv:astro-ph/0107512 [astro-ph]}
  \BibitemShut {NoStop}%
\bibitem [{\citenamefont {Carroll}(1998)}]{Carroll}%
  \BibitemOpen
  \bibfield  {author} {\bibinfo {author} {\bibfnamefont {S.~M.}\ \bibnamefont
  {Carroll}},\ }\href {\doibase 10.1103/PhysRevLett.81.3067} {\bibfield
  {journal} {\bibinfo  {journal} {Phys. Rev. Lett.}\ }\textbf {\bibinfo
  {volume} {81}},\ \bibinfo {pages} {3067} (\bibinfo {year} {1998})},\ \Eprint
  {http://arxiv.org/abs/astro-ph/9806099} {arXiv:astro-ph/9806099 [astro-ph]}
  \BibitemShut {NoStop}%
\bibitem [{\citenamefont {Dvali}\ and\ \citenamefont
  {Zaldarriaga}(2002)}]{Dvali}%
  \BibitemOpen
  \bibfield  {author} {\bibinfo {author} {\bibfnamefont {G.}~\bibnamefont
  {Dvali}}\ and\ \bibinfo {author} {\bibfnamefont {M.}~\bibnamefont
  {Zaldarriaga}},\ }\href {\doibase 10.1103/PhysRevLett.88.091303} {\bibfield
  {journal} {\bibinfo  {journal} {Phys.Rev.Lett.}\ }\textbf {\bibinfo {volume}
  {88}},\ \bibinfo {pages} {091303} (\bibinfo {year} {2002})},\ \Eprint
  {http://arxiv.org/abs/hep-ph/0108217} {arXiv:hep-ph/0108217 [hep-ph]}
  \BibitemShut {NoStop}%
\bibitem [{\citenamefont {Chiba}\ and\ \citenamefont {Kohri}(2002)}]{Chiba}%
  \BibitemOpen
  \bibfield  {author} {\bibinfo {author} {\bibfnamefont {T.}~\bibnamefont
  {Chiba}}\ and\ \bibinfo {author} {\bibfnamefont {K.}~\bibnamefont {Kohri}},\
  }\href {\doibase 10.1143/PTP.107.631} {\bibfield  {journal} {\bibinfo
  {journal} {Prog. Theor. Phys.}\ }\textbf {\bibinfo {volume} {107}},\ \bibinfo
  {pages} {631} (\bibinfo {year} {2002})},\ \Eprint
  {http://arxiv.org/abs/hep-ph/0111086} {arXiv:hep-ph/0111086 [hep-ph]}
  \BibitemShut {NoStop}%
\bibitem [{\citenamefont {Leal}\ \emph {et~al.}(2014)\citenamefont {Leal},
  \citenamefont {Martins},\ and\ \citenamefont {Ventura}}]{Leal}%
  \BibitemOpen
  \bibfield  {author} {\bibinfo {author} {\bibfnamefont {P.~M.~M.}\
  \bibnamefont {Leal}}, \bibinfo {author} {\bibfnamefont {C.~J. A.~P.}\
  \bibnamefont {Martins}}, \ and\ \bibinfo {author} {\bibfnamefont {L.~B.}\
  \bibnamefont {Ventura}},\ }\href {\doibase 10.1103/PhysRevD.90.027305}
  {\bibfield  {journal} {\bibinfo  {journal} {Phys. Rev.}\ }\textbf {\bibinfo
  {volume} {D90}},\ \bibinfo {pages} {027305} (\bibinfo {year} {2014})},\
  \Eprint {http://arxiv.org/abs/1407.4099} {arXiv:1407.4099 [astro-ph.CO]}
  \BibitemShut {NoStop}%
\bibitem [{\citenamefont {Leite}\ and\ \citenamefont
  {Martins}(2016)}]{LeiteBek}%
  \BibitemOpen
  \bibfield  {author} {\bibinfo {author} {\bibfnamefont {A.~C.~O.}\
  \bibnamefont {Leite}}\ and\ \bibinfo {author} {\bibfnamefont {C.~J. A.~P.}\
  \bibnamefont {Martins}},\ }\href {\doibase 10.1103/PhysRevD.94.023503}
  {\bibfield  {journal} {\bibinfo  {journal} {Phys. Rev.}\ }\textbf {\bibinfo
  {volume} {D94}},\ \bibinfo {pages} {023503} (\bibinfo {year} {2016})},\
  \Eprint {http://arxiv.org/abs/1607.01677} {arXiv:1607.01677 [astro-ph.CO]}
  \BibitemShut {NoStop}%
\bibitem [{\citenamefont {Olive}\ and\ \citenamefont
  {Pospelov}(2002)}]{OlivePospelov}%
  \BibitemOpen
  \bibfield  {author} {\bibinfo {author} {\bibfnamefont {K.~A.}\ \bibnamefont
  {Olive}}\ and\ \bibinfo {author} {\bibfnamefont {M.}~\bibnamefont
  {Pospelov}},\ }\href {\doibase 10.1103/PhysRevD.65.085044} {\bibfield
  {journal} {\bibinfo  {journal} {Phys. Rev.}\ }\textbf {\bibinfo {volume}
  {D65}},\ \bibinfo {pages} {085044} (\bibinfo {year} {2002})},\ \Eprint
  {http://arxiv.org/abs/hep-ph/0110377} {arXiv:hep-ph/0110377 [hep-ph]}
  \BibitemShut {NoStop}%
\bibitem [{\citenamefont {{Pepe}}\ \emph {et~al.}(2013)\citenamefont {{Pepe}},
  \citenamefont {{Cristiani}}, \citenamefont {{Rebolo}}, \citenamefont
  {{Santos}}, \citenamefont {{Dekker}}, \citenamefont {{M{\'e}gevand}},
  \citenamefont {{Zerbi}}, \citenamefont {{Cabral}}, \citenamefont {{Molaro}},
  \citenamefont {{Di Marcantonio}}, \citenamefont {{Abreu}}, \citenamefont
  {{Affolter}}, \citenamefont {{Aliverti}}, \citenamefont {{Allende Prieto}},
  \citenamefont {{Amate}}, \citenamefont {{Avila}}, \citenamefont {{Baldini}},
  \citenamefont {{Bristow}}, \citenamefont {{Broeg}}, \citenamefont {{Cirami}},
  \citenamefont {{Coelho}}, \citenamefont {{Conconi}}, \citenamefont
  {{Coretti}}, \citenamefont {{Cupani}}, \citenamefont {{D'Odorico}},
  \citenamefont {{De Caprio}}, \citenamefont {{Delabre}}, \citenamefont
  {{Dorn}}, \citenamefont {{Figueira}}, \citenamefont {{Fragoso}},
  \citenamefont {{Galeotta}}, \citenamefont {{Genolet}}, \citenamefont
  {{Gomes}}, \citenamefont {{Gonz{\'a}lez Hern{\'a}ndez}}, \citenamefont
  {{Hughes}}, \citenamefont {{Iwert}}, \citenamefont {{Kerber}}, \citenamefont
  {{Landoni}}, \citenamefont {{Lizon}}, \citenamefont {{Lovis}}, \citenamefont
  {{Maire}}, \citenamefont {{Mannetta}}, \citenamefont {{Martins}},
  \citenamefont {{Monteiro}}, \citenamefont {{Oliveira}}, \citenamefont
  {{Poretti}}, \citenamefont {{Rasilla}}, \citenamefont {{Riva}}, \citenamefont
  {{Santana Tschudi}}, \citenamefont {{Santos}}, \citenamefont {{Sosnowska}},
  \citenamefont {{Sousa}}, \citenamefont {{Span{\`o}}}, \citenamefont
  {{Tenegi}}, \citenamefont {{Toso}}, \citenamefont {{Vanzella}}, \citenamefont
  {{Viel}},\ and\ \citenamefont {{Zapatero Osorio}}}]{ESPRESSO}%
  \BibitemOpen
  \bibfield  {author} {\bibinfo {author} {\bibfnamefont {F.}~\bibnamefont
  {{Pepe}}}, \bibinfo {author} {\bibfnamefont {S.}~\bibnamefont {{Cristiani}}},
  \bibinfo {author} {\bibfnamefont {R.}~\bibnamefont {{Rebolo}}}, \bibinfo
  {author} {\bibfnamefont {N.~C.}\ \bibnamefont {{Santos}}}, \bibinfo {author}
  {\bibfnamefont {H.}~\bibnamefont {{Dekker}}}, \bibinfo {author}
  {\bibfnamefont {D.}~\bibnamefont {{M{\'e}gevand}}}, \bibinfo {author}
  {\bibfnamefont {F.~M.}\ \bibnamefont {{Zerbi}}}, \bibinfo {author}
  {\bibfnamefont {A.}~\bibnamefont {{Cabral}}}, \bibinfo {author}
  {\bibfnamefont {P.}~\bibnamefont {{Molaro}}}, \bibinfo {author}
  {\bibfnamefont {P.}~\bibnamefont {{Di Marcantonio}}}, \bibinfo {author}
  {\bibfnamefont {M.}~\bibnamefont {{Abreu}}}, \bibinfo {author} {\bibfnamefont
  {M.}~\bibnamefont {{Affolter}}}, \bibinfo {author} {\bibfnamefont
  {M.}~\bibnamefont {{Aliverti}}}, \bibinfo {author} {\bibfnamefont
  {C.}~\bibnamefont {{Allende Prieto}}}, \bibinfo {author} {\bibfnamefont
  {M.}~\bibnamefont {{Amate}}}, \bibinfo {author} {\bibfnamefont
  {G.}~\bibnamefont {{Avila}}}, \bibinfo {author} {\bibfnamefont
  {V.}~\bibnamefont {{Baldini}}}, \bibinfo {author} {\bibfnamefont
  {P.}~\bibnamefont {{Bristow}}}, \bibinfo {author} {\bibfnamefont
  {C.}~\bibnamefont {{Broeg}}}, \bibinfo {author} {\bibfnamefont
  {R.}~\bibnamefont {{Cirami}}}, \bibinfo {author} {\bibfnamefont
  {J.}~\bibnamefont {{Coelho}}}, \bibinfo {author} {\bibfnamefont
  {P.}~\bibnamefont {{Conconi}}}, \bibinfo {author} {\bibfnamefont
  {I.}~\bibnamefont {{Coretti}}}, \bibinfo {author} {\bibfnamefont
  {G.}~\bibnamefont {{Cupani}}}, \bibinfo {author} {\bibfnamefont
  {V.}~\bibnamefont {{D'Odorico}}}, \bibinfo {author} {\bibfnamefont
  {V.}~\bibnamefont {{De Caprio}}}, \bibinfo {author} {\bibfnamefont
  {B.}~\bibnamefont {{Delabre}}}, \bibinfo {author} {\bibfnamefont
  {R.}~\bibnamefont {{Dorn}}}, \bibinfo {author} {\bibfnamefont
  {P.}~\bibnamefont {{Figueira}}}, \bibinfo {author} {\bibfnamefont
  {A.}~\bibnamefont {{Fragoso}}}, \bibinfo {author} {\bibfnamefont
  {S.}~\bibnamefont {{Galeotta}}}, \bibinfo {author} {\bibfnamefont
  {L.}~\bibnamefont {{Genolet}}}, \bibinfo {author} {\bibfnamefont
  {R.}~\bibnamefont {{Gomes}}}, \bibinfo {author} {\bibfnamefont {J.~I.}\
  \bibnamefont {{Gonz{\'a}lez Hern{\'a}ndez}}}, \bibinfo {author}
  {\bibfnamefont {I.}~\bibnamefont {{Hughes}}}, \bibinfo {author}
  {\bibfnamefont {O.}~\bibnamefont {{Iwert}}}, \bibinfo {author} {\bibfnamefont
  {F.}~\bibnamefont {{Kerber}}}, \bibinfo {author} {\bibfnamefont
  {M.}~\bibnamefont {{Landoni}}}, \bibinfo {author} {\bibfnamefont {J.-L.}\
  \bibnamefont {{Lizon}}}, \bibinfo {author} {\bibfnamefont {C.}~\bibnamefont
  {{Lovis}}}, \bibinfo {author} {\bibfnamefont {C.}~\bibnamefont {{Maire}}},
  \bibinfo {author} {\bibfnamefont {M.}~\bibnamefont {{Mannetta}}}, \bibinfo
  {author} {\bibfnamefont {C.}~\bibnamefont {{Martins}}}, \bibinfo {author}
  {\bibfnamefont {M.~A.}\ \bibnamefont {{Monteiro}}}, \bibinfo {author}
  {\bibfnamefont {A.}~\bibnamefont {{Oliveira}}}, \bibinfo {author}
  {\bibfnamefont {E.}~\bibnamefont {{Poretti}}}, \bibinfo {author}
  {\bibfnamefont {J.~L.}\ \bibnamefont {{Rasilla}}}, \bibinfo {author}
  {\bibfnamefont {M.}~\bibnamefont {{Riva}}}, \bibinfo {author} {\bibfnamefont
  {S.}~\bibnamefont {{Santana Tschudi}}}, \bibinfo {author} {\bibfnamefont
  {P.}~\bibnamefont {{Santos}}}, \bibinfo {author} {\bibfnamefont
  {D.}~\bibnamefont {{Sosnowska}}}, \bibinfo {author} {\bibfnamefont
  {S.}~\bibnamefont {{Sousa}}}, \bibinfo {author} {\bibfnamefont
  {P.}~\bibnamefont {{Span{\`o}}}}, \bibinfo {author} {\bibfnamefont
  {F.}~\bibnamefont {{Tenegi}}}, \bibinfo {author} {\bibfnamefont
  {G.}~\bibnamefont {{Toso}}}, \bibinfo {author} {\bibfnamefont
  {E.}~\bibnamefont {{Vanzella}}}, \bibinfo {author} {\bibfnamefont
  {M.}~\bibnamefont {{Viel}}}, \ and\ \bibinfo {author} {\bibfnamefont {M.~R.}\
  \bibnamefont {{Zapatero Osorio}}},\ }\href@noop {} {\bibfield  {journal}
  {\bibinfo  {journal} {The Messenger}\ }\textbf {\bibinfo {volume} {153}},\
  \bibinfo {pages} {6} (\bibinfo {year} {2013})}\BibitemShut {NoStop}%
\bibitem [{\citenamefont {Liske}\ \emph {et~al.}(2014)\citenamefont {Liske}
  \emph {et~al.}}]{HIRES}%
  \BibitemOpen
  \bibfield  {author} {\bibinfo {author} {\bibfnamefont {J.}~\bibnamefont
  {Liske}} \emph {et~al.},\ }\href@noop {} {\emph {\bibinfo {title} {{Top Level
  Requirements For ELT-HIRES}}}},\ \bibinfo {type} {Tech. Rep.}\ (\bibinfo
  {institution} {Document ESO 204697 Version 1},\ \bibinfo {year}
  {2014})\BibitemShut {NoStop}%
\bibitem [{\citenamefont {Uzan}(2011)}]{Uzan}%
  \BibitemOpen
  \bibfield  {author} {\bibinfo {author} {\bibfnamefont {J.-P.}\ \bibnamefont
  {Uzan}},\ }\href {\doibase 10.12942/lrr-2011-2} {\bibfield  {journal}
  {\bibinfo  {journal} {Living Rev. Rel.}\ }\textbf {\bibinfo {volume} {14}},\
  \bibinfo {pages} {2} (\bibinfo {year} {2011})},\ \Eprint
  {http://arxiv.org/abs/1009.5514} {arXiv:1009.5514 [astro-ph.CO]} \BibitemShut
  {NoStop}%
\bibitem [{\citenamefont {Ade}\ \emph {et~al.}(2016)\citenamefont {Ade} \emph
  {et~al.}}]{Planck}%
  \BibitemOpen
  \bibfield  {author} {\bibinfo {author} {\bibfnamefont {P.~A.~R.}\
  \bibnamefont {Ade}} \emph {et~al.} (\bibinfo {collaboration} {Planck}),\
  }\href {\doibase 10.1051/0004-6361/201525830} {\bibfield  {journal} {\bibinfo
   {journal} {Astron. Astrophys.}\ }\textbf {\bibinfo {volume} {594}},\
  \bibinfo {pages} {A13} (\bibinfo {year} {2016})},\ \Eprint
  {http://arxiv.org/abs/1502.01589} {arXiv:1502.01589 [astro-ph.CO]}
  \BibitemShut {NoStop}%
\bibitem [{\citenamefont {Petrov}\ \emph {et~al.}(2006)\citenamefont {Petrov},
  \citenamefont {Nazarov}, \citenamefont {Onegin}, \citenamefont {Petrov},\
  and\ \citenamefont {Sakhnovsky}}]{Oklo}%
  \BibitemOpen
  \bibfield  {author} {\bibinfo {author} {\bibfnamefont {{\relax Yu}.~V.}\
  \bibnamefont {Petrov}}, \bibinfo {author} {\bibfnamefont {A.~I.}\
  \bibnamefont {Nazarov}}, \bibinfo {author} {\bibfnamefont {M.~S.}\
  \bibnamefont {Onegin}}, \bibinfo {author} {\bibfnamefont {V.~{\relax Yu}.}\
  \bibnamefont {Petrov}}, \ and\ \bibinfo {author} {\bibfnamefont {E.~G.}\
  \bibnamefont {Sakhnovsky}},\ }\href {\doibase 10.1103/PhysRevC.74.064610}
  {\bibfield  {journal} {\bibinfo  {journal} {Phys. Rev.}\ }\textbf {\bibinfo
  {volume} {C74}},\ \bibinfo {pages} {064610} (\bibinfo {year} {2006})},\
  \Eprint {http://arxiv.org/abs/hep-ph/0506186} {arXiv:hep-ph/0506186 [hep-ph]}
  \BibitemShut {NoStop}%
\bibitem [{\citenamefont {Rosenband}\ \emph {et~al.}(2008)\citenamefont
  {Rosenband}, \citenamefont {Hume}, \citenamefont {Schmidt}, \citenamefont
  {Chou}, \citenamefont {Brusch}, \citenamefont {Lorini}, \citenamefont
  {Oskay}, \citenamefont {Drullinger}, \citenamefont {Fortier}, \citenamefont
  {Stalnaker}, \citenamefont {Diddams}, \citenamefont {Swann}, \citenamefont
  {Newbury}, \citenamefont {Itano}, \citenamefont {Wineland},\ and\
  \citenamefont {Bergquist}}]{Rosenband}%
  \BibitemOpen
  \bibfield  {author} {\bibinfo {author} {\bibfnamefont {T.}~\bibnamefont
  {Rosenband}}, \bibinfo {author} {\bibfnamefont {D.}~\bibnamefont {Hume}},
  \bibinfo {author} {\bibfnamefont {P.}~\bibnamefont {Schmidt}}, \bibinfo
  {author} {\bibfnamefont {C.}~\bibnamefont {Chou}}, \bibinfo {author}
  {\bibfnamefont {A.}~\bibnamefont {Brusch}}, \bibinfo {author} {\bibfnamefont
  {L.}~\bibnamefont {Lorini}}, \bibinfo {author} {\bibfnamefont
  {W.}~\bibnamefont {Oskay}}, \bibinfo {author} {\bibfnamefont
  {R.}~\bibnamefont {Drullinger}}, \bibinfo {author} {\bibfnamefont
  {T.}~\bibnamefont {Fortier}}, \bibinfo {author} {\bibfnamefont
  {J.}~\bibnamefont {Stalnaker}}, \bibinfo {author} {\bibfnamefont
  {S.}~\bibnamefont {Diddams}}, \bibinfo {author} {\bibfnamefont
  {W.}~\bibnamefont {Swann}}, \bibinfo {author} {\bibfnamefont
  {N.}~\bibnamefont {Newbury}}, \bibinfo {author} {\bibfnamefont
  {W.}~\bibnamefont {Itano}}, \bibinfo {author} {\bibfnamefont
  {D.}~\bibnamefont {Wineland}}, \ and\ \bibinfo {author} {\bibfnamefont
  {J.}~\bibnamefont {Bergquist}},\ }\href {\doibase 10.1126/science.1154622}
  {\bibfield  {journal} {\bibinfo  {journal} {Science}\ }\textbf {\bibinfo
  {volume} {319}},\ \bibinfo {pages} {1808} (\bibinfo {year}
  {2008})}\BibitemShut {NoStop}%
\bibitem [{\citenamefont {Murphy}\ \emph {et~al.}(2016)\citenamefont {Murphy},
  \citenamefont {Malec},\ and\ \citenamefont {Prochaska}}]{MalecNew}%
  \BibitemOpen
  \bibfield  {author} {\bibinfo {author} {\bibfnamefont {M.~T.}\ \bibnamefont
  {Murphy}}, \bibinfo {author} {\bibfnamefont {A.~L.}\ \bibnamefont {Malec}}, \
  and\ \bibinfo {author} {\bibfnamefont {J.~X.}\ \bibnamefont {Prochaska}},\
  }\href {\doibase 10.1093/mnras/stw1482} {\bibfield  {journal} {\bibinfo
  {journal} {MNRAS}\ }\textbf {\bibinfo {volume} {461}},\ \bibinfo {pages}
  {2461} (\bibinfo {year} {2016})},\ \bibinfo {note} {[Erratum: MNRAS
  464,2609(2017)]},\ \Eprint {http://arxiv.org/abs/1606.06293}
  {arXiv:1606.06293 [astro-ph.CO]} \BibitemShut {NoStop}%
\bibitem [{\citenamefont {Songaila}\ and\ \citenamefont
  {Cowie}(2014)}]{Songaila}%
  \BibitemOpen
  \bibfield  {author} {\bibinfo {author} {\bibfnamefont {A.}~\bibnamefont
  {Songaila}}\ and\ \bibinfo {author} {\bibfnamefont {L.~L.}\ \bibnamefont
  {Cowie}},\ }\href {\doibase 10.1088/0004-637X/793/2/103} {\bibfield
  {journal} {\bibinfo  {journal} {Astrophys.J.}\ }\textbf {\bibinfo {volume}
  {793}},\ \bibinfo {pages} {103} (\bibinfo {year} {2014})},\ \Eprint
  {http://arxiv.org/abs/1406.3628} {arXiv:1406.3628 [astro-ph.CO]} \BibitemShut
  {NoStop}%
\bibitem [{\citenamefont {{Evans}}\ \emph {et~al.}(2014)\citenamefont
  {{Evans}}, \citenamefont {{Murphy}}, \citenamefont {{Whitmore}},
  \citenamefont {{Misawa}}, \citenamefont {{Centurion}}, \citenamefont
  {{D'Odorico}}, \citenamefont {{Lopez}}, \citenamefont {{Martins}},
  \citenamefont {{Molaro}}, \citenamefont {{Petitjean}}, \citenamefont
  {{Rahmani}}, \citenamefont {{Srianand}},\ and\ \citenamefont
  {{Wendt}}}]{LP3}%
  \BibitemOpen
  \bibfield  {author} {\bibinfo {author} {\bibfnamefont {T.~M.}\ \bibnamefont
  {{Evans}}}, \bibinfo {author} {\bibfnamefont {M.~T.}\ \bibnamefont
  {{Murphy}}}, \bibinfo {author} {\bibfnamefont {J.~B.}\ \bibnamefont
  {{Whitmore}}}, \bibinfo {author} {\bibfnamefont {T.}~\bibnamefont
  {{Misawa}}}, \bibinfo {author} {\bibfnamefont {M.}~\bibnamefont
  {{Centurion}}}, \bibinfo {author} {\bibfnamefont {S.}~\bibnamefont
  {{D'Odorico}}}, \bibinfo {author} {\bibfnamefont {S.}~\bibnamefont
  {{Lopez}}}, \bibinfo {author} {\bibfnamefont {C.~J.~A.~P.}\ \bibnamefont
  {{Martins}}}, \bibinfo {author} {\bibfnamefont {P.}~\bibnamefont {{Molaro}}},
  \bibinfo {author} {\bibfnamefont {P.}~\bibnamefont {{Petitjean}}}, \bibinfo
  {author} {\bibfnamefont {H.}~\bibnamefont {{Rahmani}}}, \bibinfo {author}
  {\bibfnamefont {R.}~\bibnamefont {{Srianand}}}, \ and\ \bibinfo {author}
  {\bibfnamefont {M.}~\bibnamefont {{Wendt}}},\ }\href {\doibase
  10.1093/mnras/stu1754} {\bibfield  {journal} {\bibinfo  {journal}
  {Mon.Not.Roy.Astron.Soc.}\ }\textbf {\bibinfo {volume} {445}},\ \bibinfo
  {pages} {128} (\bibinfo {year} {2014})},\ \Eprint
  {http://arxiv.org/abs/1409.1923} {arXiv:1409.1923} \BibitemShut {NoStop}%
\bibitem [{\citenamefont {{Kotu{\v s}}}\ \emph {et~al.}(2017)\citenamefont
  {{Kotu{\v s}}}, \citenamefont {{Murphy}},\ and\ \citenamefont
  {{Carswell}}}]{Kotus}%
  \BibitemOpen
  \bibfield  {author} {\bibinfo {author} {\bibfnamefont {S.~M.}\ \bibnamefont
  {{Kotu{\v s}}}}, \bibinfo {author} {\bibfnamefont {M.~T.}\ \bibnamefont
  {{Murphy}}}, \ and\ \bibinfo {author} {\bibfnamefont {R.~F.}\ \bibnamefont
  {{Carswell}}},\ }\href {\doibase 10.1093/mnras/stw2543} {\bibfield  {journal}
  {\bibinfo  {journal} {MNRAS}\ }\textbf {\bibinfo {volume} {464}},\ \bibinfo
  {pages} {3679} (\bibinfo {year} {2017})},\ \Eprint
  {http://arxiv.org/abs/1609.03860} {arXiv:1609.03860} \BibitemShut {NoStop}%
\bibitem [{\citenamefont {{Agafonova}}\ \emph {et~al.}(2011)\citenamefont
  {{Agafonova}}, \citenamefont {{Molaro}}, \citenamefont {{Levshakov}},\ and\
  \citenamefont {{Hou}}}]{alphaAgafonova}%
  \BibitemOpen
  \bibfield  {author} {\bibinfo {author} {\bibfnamefont {I.~I.}\ \bibnamefont
  {{Agafonova}}}, \bibinfo {author} {\bibfnamefont {P.}~\bibnamefont
  {{Molaro}}}, \bibinfo {author} {\bibfnamefont {S.~A.}\ \bibnamefont
  {{Levshakov}}}, \ and\ \bibinfo {author} {\bibfnamefont {J.~L.}\ \bibnamefont
  {{Hou}}},\ }\href {\doibase 10.1051/0004-6361/201016194} {\bibfield
  {journal} {\bibinfo  {journal} {A. \& A.}\ }\textbf {\bibinfo {volume}
  {529}},\ \bibinfo {eid} {A28} (\bibinfo {year} {2011})},\ \Eprint
  {http://arxiv.org/abs/1102.2967} {arXiv:1102.2967 [astro-ph.CO]} \BibitemShut
  {NoStop}%
\bibitem [{\citenamefont {Molaro}\ \emph {et~al.}(2013)\citenamefont {Molaro},
  \citenamefont {Centurion}, \citenamefont {Whitmore}, \citenamefont {Evans},
  \citenamefont {Murphy} \emph {et~al.}}]{LP1}%
  \BibitemOpen
  \bibfield  {author} {\bibinfo {author} {\bibfnamefont {P.}~\bibnamefont
  {Molaro}}, \bibinfo {author} {\bibfnamefont {M.}~\bibnamefont {Centurion}},
  \bibinfo {author} {\bibfnamefont {J.}~\bibnamefont {Whitmore}}, \bibinfo
  {author} {\bibfnamefont {T.}~\bibnamefont {Evans}}, \bibinfo {author}
  {\bibfnamefont {M.}~\bibnamefont {Murphy}},  \emph {et~al.},\ }\href
  {\doibase 10.1051/0004-6361/201321351} {\bibfield  {journal} {\bibinfo
  {journal} {Astron.Astrophys.}\ }\textbf {\bibinfo {volume} {555}},\ \bibinfo
  {pages} {A68} (\bibinfo {year} {2013})},\ \Eprint
  {http://arxiv.org/abs/1305.1884} {arXiv:1305.1884 [astro-ph.CO]} \BibitemShut
  {NoStop}%
\bibitem [{\citenamefont {{Bainbridge}}\ and\ \citenamefont
  {{Webb}}(2017)}]{Bainbridge}%
  \BibitemOpen
  \bibfield  {author} {\bibinfo {author} {\bibfnamefont {M.~B.}\ \bibnamefont
  {{Bainbridge}}}\ and\ \bibinfo {author} {\bibfnamefont {J.~K.}\ \bibnamefont
  {{Webb}}},\ }\href {\doibase 10.1093/mnras/stx179} {\bibfield  {journal}
  {\bibinfo  {journal} {Mon. Not. Roy. Astron. Soc.}\ }\textbf {\bibinfo
  {volume} {468}},\ \bibinfo {pages} {1639} (\bibinfo {year} {2017})},\ \Eprint
  {http://arxiv.org/abs/1606.07393} {arXiv:1606.07393 [astro-ph.IM]}
  \BibitemShut {NoStop}%
\bibitem [{\citenamefont {Jones}\ \emph {et~al.}(2017)\citenamefont {Jones}
  \emph {et~al.}}]{Jones}%
  \BibitemOpen
  \bibfield  {author} {\bibinfo {author} {\bibfnamefont {D.~O.}\ \bibnamefont
  {Jones}} \emph {et~al.},\ }\href@noop {} {\  (\bibinfo {year} {2017})},\
  \Eprint {http://arxiv.org/abs/1710.00846} {arXiv:1710.00846 [astro-ph.CO]}
  \BibitemShut {NoStop}%
\bibitem [{\citenamefont {Suzuki}\ \emph {et~al.}(2012)\citenamefont {Suzuki},
  \citenamefont {Rubin}, \citenamefont {Lidman}, \citenamefont {Aldering},
  \citenamefont {Amanullah} \emph {et~al.}}]{Union}%
  \BibitemOpen
  \bibfield  {author} {\bibinfo {author} {\bibfnamefont {N.}~\bibnamefont
  {Suzuki}}, \bibinfo {author} {\bibfnamefont {D.}~\bibnamefont {Rubin}},
  \bibinfo {author} {\bibfnamefont {C.}~\bibnamefont {Lidman}}, \bibinfo
  {author} {\bibfnamefont {G.}~\bibnamefont {Aldering}}, \bibinfo {author}
  {\bibfnamefont {R.}~\bibnamefont {Amanullah}},  \emph {et~al.},\ }\href
  {\doibase 10.1088/0004-637X/746/1/85} {\bibfield  {journal} {\bibinfo
  {journal} {Astrophys.J.}\ }\textbf {\bibinfo {volume} {746}},\ \bibinfo
  {pages} {85} (\bibinfo {year} {2012})},\ \Eprint
  {http://arxiv.org/abs/1105.3470} {arXiv:1105.3470 [astro-ph.CO]} \BibitemShut
  {NoStop}%
\bibitem [{\citenamefont {Farooq}\ \emph {et~al.}(2017)\citenamefont {Farooq},
  \citenamefont {Madiyar}, \citenamefont {Crandall},\ and\ \citenamefont
  {Ratra}}]{Farooq}%
  \BibitemOpen
  \bibfield  {author} {\bibinfo {author} {\bibfnamefont {O.}~\bibnamefont
  {Farooq}}, \bibinfo {author} {\bibfnamefont {F.~R.}\ \bibnamefont {Madiyar}},
  \bibinfo {author} {\bibfnamefont {S.}~\bibnamefont {Crandall}}, \ and\
  \bibinfo {author} {\bibfnamefont {B.}~\bibnamefont {Ratra}},\ }\href
  {\doibase 10.3847/1538-4357/835/1/26} {\bibfield  {journal} {\bibinfo
  {journal} {Astrophys. J.}\ }\textbf {\bibinfo {volume} {835}},\ \bibinfo
  {pages} {26} (\bibinfo {year} {2017})},\ \Eprint
  {http://arxiv.org/abs/1607.03537} {arXiv:1607.03537 [astro-ph.CO]}
  \BibitemShut {NoStop}%
\bibitem [{\citenamefont {Leite}(2015)}]{MSC}%
  \BibitemOpen
  \bibfield  {author} {\bibinfo {author} {\bibfnamefont {A.~C.~O.}\
  \bibnamefont {Leite}},\ }\emph {\bibinfo {title} {{Optimization of ESPRESSO
  Fundamental Physics Tests}}},\ \href@noop {} {Master's thesis},\ \bibinfo
  {school} {University of Porto} (\bibinfo {year} {2015})\BibitemShut {NoStop}%
\bibitem [{\citenamefont {Leite}\ \emph {et~al.}(2016)\citenamefont {Leite},
  \citenamefont {Martins}, \citenamefont {Molaro}, \citenamefont {Corre},\ and\
  \citenamefont {Cristiani}}]{LMMCC}%
  \BibitemOpen
  \bibfield  {author} {\bibinfo {author} {\bibfnamefont {A.~C.~O.}\
  \bibnamefont {Leite}}, \bibinfo {author} {\bibfnamefont {C.~J. A.~P.}\
  \bibnamefont {Martins}}, \bibinfo {author} {\bibfnamefont {P.}~\bibnamefont
  {Molaro}}, \bibinfo {author} {\bibfnamefont {D.}~\bibnamefont {Corre}}, \
  and\ \bibinfo {author} {\bibfnamefont {S.}~\bibnamefont {Cristiani}},\ }\href
  {\doibase 10.1103/PhysRevD.94.123512} {\bibfield  {journal} {\bibinfo
  {journal} {Phys. Rev.}\ }\textbf {\bibinfo {volume} {D94}},\ \bibinfo {pages}
  {123512} (\bibinfo {year} {2016})},\ \Eprint
  {http://arxiv.org/abs/1612.05284} {arXiv:1612.05284 [astro-ph.CO]}
  \BibitemShut {NoStop}%
\bibitem [{\citenamefont {Levshakov}\ \emph {et~al.}(2013)\citenamefont
  {Levshakov}, \citenamefont {Reimers}, \citenamefont {Henkel}, \citenamefont
  {Winkel}, \citenamefont {Mignano}, \citenamefont {Centurion},\ and\
  \citenamefont {Molaro}}]{Levshakov}%
  \BibitemOpen
  \bibfield  {author} {\bibinfo {author} {\bibfnamefont {S.~A.}\ \bibnamefont
  {Levshakov}}, \bibinfo {author} {\bibfnamefont {D.}~\bibnamefont {Reimers}},
  \bibinfo {author} {\bibfnamefont {C.}~\bibnamefont {Henkel}}, \bibinfo
  {author} {\bibfnamefont {B.}~\bibnamefont {Winkel}}, \bibinfo {author}
  {\bibfnamefont {A.}~\bibnamefont {Mignano}}, \bibinfo {author} {\bibfnamefont
  {M.}~\bibnamefont {Centurion}}, \ and\ \bibinfo {author} {\bibfnamefont
  {P.}~\bibnamefont {Molaro}},\ }\href {\doibase 10.1051/0004-6361/201322535}
  {\bibfield  {journal} {\bibinfo  {journal} {Astron. Astrophys.}\ }\textbf
  {\bibinfo {volume} {559}},\ \bibinfo {pages} {A91} (\bibinfo {year}
  {2013})},\ \Eprint {http://arxiv.org/abs/1310.1850} {arXiv:1310.1850
  [astro-ph.GA]} \BibitemShut {NoStop}%
\end{thebibliography}%
\end{document}